\documentclass[preprint,prd,aps,showpacs,preprintnumbers,amsmath,amssymb]{revtex4-1}
\usepackage{graphics,epsfig,subfigure}
\usepackage{diagbox}
\usepackage[usenames]{color}
\usepackage{color}
\usepackage{graphicx}
\usepackage{amsfonts}
\usepackage{indentfirst}
\usepackage{booktabs}
\usepackage{hyperref}
\hypersetup{hypertex=ture,backref=true,colorlinks=ture,linkcolor=blue,anchorcolor=blue,citecolor=blue}
\usepackage{float}
\usepackage{multirow}
\usepackage[section]{placeins}

\begin{document}

\title{\Large \bf Building a black hole-wormhole-black hole combination}
\author{Xin Su}
\author{Chen-Hao Hao}
\author{Ji-Rong Ren}
\author{Yong-Qiang Wang\footnote{yqwang@lzu.edu.cn, corresponding author}}
\affiliation{ $^{1}$Lanzhou Center for Theoretical Physics, Key Laboratory of Theoretical Physics of Gansu Province,
	School of Physical Science and Technology, Lanzhou University, Lanzhou 730000, China\\
	$^{2}$Institute of Theoretical Physics $\&$ Research Center of Gravitation, Lanzhou University, Lanzhou 730000, China}

\begin{abstract}
In this paper, we present the spherically symmetric Proca star in the presence of a phantom field and obtain a traversable wormhole solution for non-trivial topological spacetime. Using numerical methods, symmetric solutions and asymmetric solutions are obtained in two asymptotically flat regions. We find that when changing the throat size $r_{0}$, both the mass $M$ and the Noether charge $Q$ no longer have the spiral characteristics of an independent Proca star, furthermore, the asymmetric solution can be turned into the symmetric solution at some frequency $\omega$ in certain $r_{0}$. In particular, we find that when the frequency takes a certain value, for each solution, there is an extremely approximate black hole solution, and there is even a case where an event horizon appears on both sides of the wormhole throat. 
\end{abstract}

\maketitle

\section{INTRODUCTION}\label{Sec1}

The minimal coupling of Einstein's gravitational field and scalar field will result in a local particle-like solution, which is called the Boson star \cite{Wheeler:1955zz,Kaup:1968zz,Ruffini:1969qy}. It was originally proposed to realize J. A. Wheeler's geon idea {\cite{Wheeler:1955zz,Power:1957zz}. Similar to Boson stars, gravity can also couple to fields of non-zero spin. In 2015, research by R. Brito et al. showed that the coupling of gravity and the Proca field with spin 1 can obtain a static solution, which is called the Proca star \cite{Brito:2015pxa}. After that, I. Salazarlandea and F. Garciaka constructed a model of a charged Proca star \cite{SalazarLandea:2016bys}. In different spacetime backgrounds, R. Brito investigated Proca stars in Anti-de Sitter (AdS) spacetime \cite{Duarte:2016lig}, while Y. Brihaye et al. explored the counterpart of Proca stars in flat spacetime, known as Proca Q-balls\cite{Loginov:2015rya,Brihaye:2017inn,Heeck:2021bce,Dzhunushaliev:2021tlm,Herdeiro:2023lze}. In addition, rotating Proca stars and Proca stars with self-interaction have also been studied by researchers \cite{Herdeiro:2019mbz,Aoki:2022woy,Aoki:2022mdn}. Actually, the properties of Proca stars and Boson stars have many similarities, such as they both have conserved charge and time dependence, but Proca stars have better dynamic stability \cite{Sanchis-Gual:2019ljs}. In recent years, with the development of astrophysics, Proca stars have been considered as one of the candidates for dark matter. It has also been widely used in black hole shadow simulation  \cite{Herdeiro:2021lwl} and gravitational wave signal analysis \cite{Sanchis-Gual:2018oui,CalderonBustillo:2020fyi,Sanchis-Gual:2022mkk}. In addition, there are some interesting studies on Proca stars listed in \cite{Shen:2016acv,Minamitsuji:2017pdr,Santos:2020pmh,Rosa:2022tfv,Ma:2023bhb}.

In the context of general relativity, in addition to forming the astrophysical compact objects such as Boson stars and Proca stars, wormholes as a peculiar space-time structure can also be formed which is attracting more and more attention. The concept of a wormhole was first introduced in 1916 by the German physicist L. Flamm. It was not until 1935 that A. Einstein and N. Rosen proposed the concept of the Einstein-Rosen bridge and supplemented the wormhole theory \cite{Einstein:1935tc}. Wormhole is a tunnel that can connect two separate parts of the same world or universe, allowing for interstellar travel. However, for an observer trying to pass through, the wormhole opens up and closes too quickly for even a photon to get through \cite{Kruskal:1959vx,Fuller:1962zza}. Until H. G. Ellis \cite{Ellis:1973yv} and K. A. Bronnikov \cite{Bronnikov:1973fh} conducted the earliest research on traversable wormholes. Later, the work of M. Morris and K. Thorne sparked a growing interest in the solution of traversable wormholes \cite{Morris:1988cz}. Their research showed that exotic matter with negative energy density can maintain the opening of wormholes and form traversable wormholes, but this violates the Null Energy Condition (NEC) \cite{Visser:1989kh}. Several types of such traversable wormholes have actually been discovered in \cite{Ellis:1979bh,Kodama:1978dw} , and these traversable wormhole models have chosen phantom scalar fields with dynamical terms of opposite signs as potential candidates for this exotic matter. In addition, there are also traversable wormholes that are not constructed with exotic matter\cite{Blazquez-Salcedo:2020czn,Konoplya:2021hsm,Wang:2022aze,Bronnikov:2002rn,Lobo:2009ip,Rahaman:2013qza,Rani:2016xpa,Sahoo:2017ual,Klinkhamer:2022rsj,Klinkhamer:2023sau,Klinkhamer:2023avf,Wang:2023rfq}.

Previously, research on compact stars such as Boson stars mainly focused on space-time with trivial topology. V. Dzhunushaliev first connected stars and wormhole spacetime by considering mixed systems with wormholes in the core\cite{Dzhunushaliev:2011xx}. In \cite{Dzhunushaliev:2014bya,Hoffmann:2017jfs,Yue:2023ela,Ding:2023syj}, Boson stars under wormhole space-time were studied. The Boson matter that makes up this system consists of a scalar field or a complex scalar field with self-interaction, which can exhibit symmetric or asymmetric distribution on two asymptotically flat regions. In a recent study \cite{Hao:2023igi}, the authors investigated Dirac stars within the context of wormholes. Remarkably, they unveiled an extremely approximate black hole solution within the framework of the static wormhole model. In fact, wormholes are closely related to black holes. To some extent, wormholes can mimic the observational features of black holes \cite{Damour:2007ap}. In addition, Maldacena and Susskind first proposed the idea of ER=EPR, believing that two black holes can be connected via wormholes to form a pair of entangled states \cite{Maldacena:2013xja}. However, no such static wormhole model with black holes has been found so far. In our work, we chose the phantom scalar field as the exotic matter, constructed a proca star model with a wormhole in non-trivial topological space-time, studied the symmetric and asymmetric solutions of the proca field in this hybrid system, and described it using numerical methods. Not only do extremely approximate black holes appear in the static wormhole, but for the first time, two event horizons appear on both sides of the throat. This can be viewed as a black hole-wormhole-black hole combination. This may provide a specific model for the ER=EPR theory.

This paper is structured as follows. In Sec.~\ref{sec2}, we use the phantom field as an exotic matter to construct a Proca star with a wormhole in the center and give the expressions of the ADM mass and the Noether charge that we are interested in. In Sec.~\ref{sec3}, we mainly give the boundary conditions required for the solution. We mainly classify and summarize our numerical results in Sec.~\ref{sec4}. The last section is a summary and outlook of the full text.\raggedbottom
 
\section{THE MODEL}\label{sec2}

\subsection{Action}

We consider the Proca field and the phantom scalar field with minimally coupling to the Einstein tensor,  the corresponding Einstein-Hilbert action is
\begin{equation}\label{action}
  S=\int\sqrt{-g}d^4x\left(\frac{R}{2\kappa}+\mathcal{L}_{p}+\mathcal{L}_{m}\right),
\end{equation}
where $\kappa$ represents the coupling constant, $R$ is the Riccci scalar. The term $\mathcal{L}_{p}$ and $\mathcal{L}_{m}$ are the Lagrangians of the phantom field and the proca field respectively, where the phantom field has the negative kinetic term. Their specific forms are as follows
\begin{equation}
\mathcal{L}_{(m)}=-\frac{1}{4} \mathcal{F}_{\alpha \beta} \overline{\mathcal{F}}^{\alpha \beta}-\frac{1}{2} \mu^{2} \mathcal{A}_{\alpha} \overline{\mathcal{A}}^{\alpha},
\end{equation}
\begin{equation}
\mathcal{L}_{p}=\nabla_{a} \Phi \nabla^{a} \Phi.
\end{equation}
Here $\mathcal{A}$ and  $\Phi$ represent the Proca field and the phantom field, respectively. $ \overline{\mathcal{A}}$ is the conjugate term, $\mathcal{F}=d\mathcal{A}$. $\mu$ is the mass of the Proca field.

Varying the action (1) with respect to the metric tensor leads to the Einstein equations of the mixed system  
\begin{equation}
R_{\mu \nu}-\frac{1}{2} g_{\mu \nu} R=\kappa T_{\mu \nu},
\end{equation}
and stress-energy tensor
\begin{equation}
T_{\mu \nu}=g_{\mu \nu}\left(\mathcal{L}_{m}+\mathcal{L}_{p}\right)-2 \frac{\partial\left(\mathcal{L}_{m}+\mathcal{L}_{p}\right)}{\partial g^{\mu \nu}}.
\end{equation}

The matter field equations can be obtained by varying with respect to the Proca field and phantom field
\begin{equation}
\nabla_{\alpha} \mathcal{F}^{\alpha \beta}-\mu^{2} \mathcal{A}^{\beta}=0,
\end{equation}
\begin{equation}
\nabla^{2}\Phi=0.
\end{equation}
In addition, the Proca field satisfies the following Lorentz condition, which is very helpful for simplifying the equation:
\begin{equation}
\nabla_{\alpha} \mathcal{A}^{\alpha}=0.
\end{equation}

\subsection{Ansatz}
We consider the static spherical symmetry solution with a wormhole, and the ansatz \cite{Hoffmann:2017jfs} is as follows
\begin{equation}  \label{line_element1}
 ds^2 = -e^{A}  dt^2 +B e^{-A}   \left[ d r^2 + h (d \theta^2+\sin^2 \theta d\varphi^2)   \right]\,,
\end{equation}
where $A$ and $B$ are both functions of the radial coordinate $r$, $h=r^{2}+r_{0}^{2}$, $r_{0}$ is the throat siza. The range of $r$ is from negative infinity to positive infinity. When $r$ approaches positive and negative infinity, it corresponds to two asymptotically flat spacetimes.

Furthermore, for the static spherically symmetric system, we use the following ansatz of the Proca field and the phantom field
\begin{equation}
\mathcal{A}=[F(r)dt+iG(r)dr]e^{-i\omega t},
\end{equation}
\begin{equation}
\Phi=\phi(r).
\end{equation}
Here $\phi(r)$, $F(r)$ and $G(r)$ are functions of the coordinate $r$. $\omega$ denotes the frequency. The phantom field is independent of the time coordinate $t$.

\subsection{Einstein field equation}
Substituting the above ansatz (8) - (10) into Einstein equation (4), we get the following system of equations
\begin{equation}
\begin{split}
&2h^2\kappa e^{-A}{(F^\prime-\omega G)}^2B^2+2e^{-2A}h^2\kappa\mu^2F^2B^3-3h^2{B^\prime}^2+4B^2{r_0}^2+\\&hB^2(-8rA^\prime+h(2\kappa\mu^2G^2+{A^\prime}^2-2\kappa{\phi^\prime}^2-4A^{\prime\prime}))+2hBe^{-A}(B^\prime(4r-hA^\prime)+2hB^{\prime\prime}))=0,
\end{split}
\end{equation}

\begin{equation}
\begin{split}
&2 e^{-2A} \kappa \mu^{2} F^{2} B^{3}+B^{2}\left(2 \kappa e^{-A}\left(-\omega G+F^{\prime}\right)^{2}-2 \kappa \mu^{2} G^{2}-A^{\prime 2}+2 \kappa \phi^{\prime 2}\right)+ B^{\prime 2}+\frac{4 r B B^{\prime}}{h}\\&-\frac{4 r_{0}^{2} B^{2}}{h^{2}}=0,
\end{split}
\end{equation}

\begin{equation}
\begin{split}
&-2 e^{-2A} \kappa \mu^{2} F B^{3}-2  B^{\prime 2}+B^{2}\left(\frac{4 r_{0}^{2}}{h^{2}}-2 \kappa e^{-A}\left(-\omega G+F^{\prime}\right)^{2}+2 \kappa \mu^{2} G^{2}+A^{\prime 2}-2 \kappa \phi^{\prime 2}\right)\\&+2 B\left(\frac{r B^{\prime}}{h}+B^{\prime \prime}\right)=0 .
\end{split}
\end{equation}
arising from the $tt$, $rr$, $\theta\theta$ components, respectively.

Variation of the action with respect to the Proca field and to the phantom field leads to the equations
\begin{equation}
(-\frac{e^A\mu^2}{\omega}+\omega)G-F^\prime=0,
\end{equation}
\begin{equation}
-\frac{4rB}{r^2+{r_{0}}^2}-\frac{2B(e^{-2A}\omega FB+G^\prime)}{G}-B^\prime=0,
\end{equation}
\begin{equation}
\left(h \sqrt{B} \phi^{\prime}\right)^{\prime}=0.
\end{equation}
Integrating the equation (17) leads to the following equation
\begin{equation}
\phi^\prime=\frac{\sqrt D}{h\sqrt B}.
\end{equation}
The constant $D$ is the scalar charge of the phantom field. Substituting equation (18) into Einstein's equation (13) can eliminate the $\phi^\prime$ term of the phantom field, and obtain the specific expression of $D$ as follows
\begin{equation}
\begin{split}
D=&\frac{1}{2\kappa B}e^{-2A}(2h^2\kappa\mu^2F^2B^3+e^AB^2(4e^A{r_{0}}^2+2h^2\kappa(e^A\mu^2-\omega^2)G^2+e^Ah^2{A^\prime}^2+\\&4h^2\kappa\omega GF^\prime-2r^4\kappa{F^\prime}^2-4r^2{r_{0}}^2\kappa{F^\prime}^2-2{r_{0}}^4\kappa{F^\prime}^2)-4e^{2A}rhBB^\prime-e^{2A}h^2{B^\prime}^2).
\end{split}
\end{equation}

In order to solve the metric function, we need to add Eq. (12) to Eq. (13), and Eq. (13) to Eq. (14) to eliminate $\phi^{\prime{2}}$ terms, the final Einstein equations can be converted into the following form to solve
\begin{equation}
\frac{1}{2}(\frac{6r}{h}-A^\prime)B^\prime-\frac{{B^{\prime{2}}}}{2B}+B(-\frac{2rA^\prime}{h}+e^{-A}\kappa{(-\omega G+F^\prime)}^2-A^{\prime\prime})+{B}^{\prime\prime}=0,
\end{equation}
\begin{equation}
-2e^{-2A}\kappa\mu^2F^2B^2+\frac{3rB^\prime}{h}-\frac{B^{\prime2}}{2B}+B^{\prime\prime}=0.
\end{equation}
Together with equations (15) and (16), we obtain a system of ODEs that can be solved numerically.

\subsection{Mass and charge}

The ADM mass can be obtained directly from the asymptotic expression of the metric component $g_{tt}$, which works for both symmetric and asymmetric solutions.
\begin{equation}
g_{tt}\rightarrow -1+\frac{2M}{R}.
\end{equation}

The action has global U(1) invariance under the transformation $A\to e^{i\alpha}A$ with a constant $\alpha$. This means that there is a conserved current 
\begin{equation}
\ j^{\alpha}=\frac{i}{2}\left[\overline{\mathcal{F}}^{\alpha \beta} \mathcal{A}_{\beta}-\mathcal{F}^{\alpha \beta} \overline{\mathcal{A}}_{\beta}\right],j^\alpha_;\alpha=0.
\end{equation}
By integrating the time component of the current, the Noether charge of the Proca field can be read
\begin{equation}
Q=-\int j^t\left | g \right | ^{1/2}drd\theta d \varphi.
\end{equation}
Noether charge $Q$ corresponds to the number of proca particles. It should be noted that formula (24) only applies to symmetric solutions, for asymmetric solutions, the calculation of $Q$ will become complicated. Since the choice of the inner boundary is ambiguous when the usual integral over the time component of the conserved current is performed, as discussed in \cite{Hoffmann:2017jfs}. The specific method introduces a coupling to a fictitious electromagnetic field, such that the $Q$ can be read off asymptotically as the electric charge associated with the gauge field, which refers to article \cite{Hoffmann:2018oml}. The specific methods are as follows

The fictitious electromagnetic field is determined from
\begin{equation}
\partial_{\mu}\left(\sqrt{-g} g^{\mu \sigma} g^{\nu \rho} F_{\sigma \rho}\right)=j^{\nu} \sqrt{-g},
\end{equation}
where$F_{\sigma \rho }=\partial _{\sigma} A_{\rho }- \partial _{\rho} A_{\sigma }$ is the field strength of the gauge potential $A_{\mu}$, and $j_{\nu}$ is the current as the form (23). The ansatz for the gauge potential we choose
\begin{equation}
A_{\mu}dx_{\mu}=a_{0}(r)dt.
\end{equation}
Subtitution in the (25) yields the function $a_{0}(r)$ to be solved. The boundary conditions we take are $a_{0}(\pm \infty )=0$.
The charges $Q_{\pm}$ are now determined from teh asymptotic behaviour of $a_{0}(r)$
\begin{equation}
a_{0}\to \pm \frac{Q_{\pm }}{r} ,\ r\to \pm \infty .
\end{equation}

\section{BOUNDARY CONDITIONS}\label{sec3}
We need eight boundary conditions about $F$, $G$, $A$, and $B$ to solve the coupled ODEs in the asymptotically flat space-time. It should be emphasized that for the Boson star of wormhole spacetime, the boson field function is symmetric or antisymmetric, and different boundary conditions need to be applied at $x=0$ to distinguish two types of solutions. However in our model, there are two field functions whose symmetry is not consistent, so we no longer apply additional boundary conditions at $x=0$.

In order to prevent energy divergence, the Proca field function needs to satisfy
\begin{equation}
F(\pm\infty )=0,\ G(\pm\infty )=0.
\end{equation}
When $r$ tends to infinity, the solution is an asymptotically flat space-time, so the metric function satisfies
\begin{equation}
A(\pm\infty )=0,\ B(\pm\infty )=1.
\end{equation}
These boundary conditions are set based on symmetric solutions, but asymmetric solutions also satisfy naturally.

\section{NUMERICAL RESULTS}\label{sec4}
Here we study two types of symmetric solutions and one asymmetric solution of the system and discuss the connection between these two configurations. For each case, we change the value of $r_{0}$ to analyze the changes in the solution and its corresponding ADM mass $M$ and Noether charge $Q$. Because we are considering the wormhole model that connects two asymptotically flat spacetime, the mass distribution on either side of the wormhole is denoted as $M_{+}$ and $M_{-}$, similarly, the distribution of Noether charge is denoted as $Q_{+}$ and $Q_{-}$.

For the numerical calculations, we compactify the coordinates by introducing a new radial coordinate $x$.
\begin{equation}
r=\tan \left(\frac{\pi}{2} x\right ),
\end{equation}
$r\in(-\infty,+\infty)$ corresponding to $x\in(-1,+1)$. We use the finite element algorithm to numerically solve the ODE. Discretize the region $-1\le x\le 1$ into a grid of 500-2000 points. In addition, the relative error should be less than $10^{-4}$ to ensure the correctness of the results. The parameter $\kappa$ is fixed to 2.

\subsection{Symmetric type \uppercase\expandafter{\romannumeral1} }

\begin{figure}
  \begin{center}
\subfigure{\includegraphics[width=0.49\textwidth]{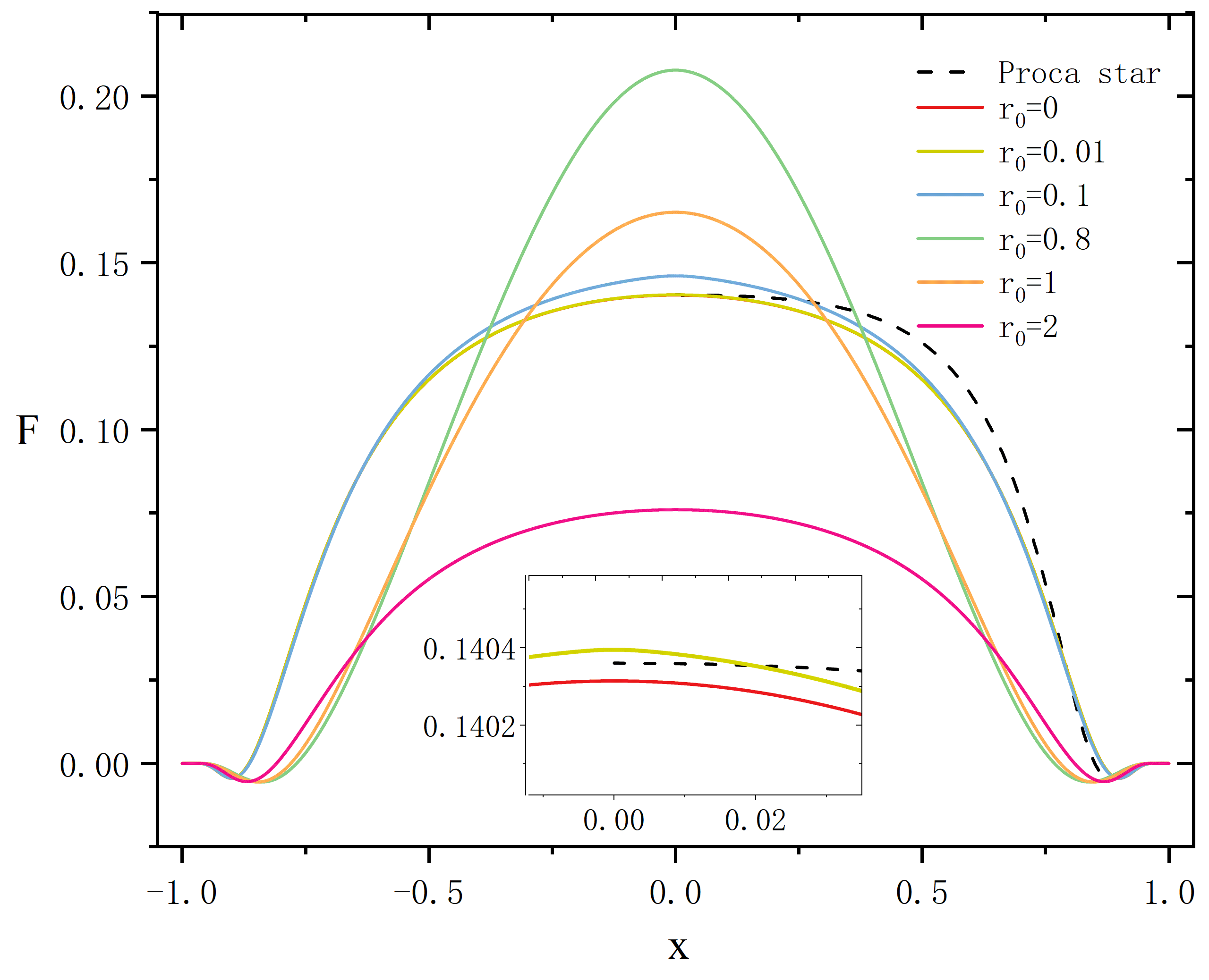}}
\subfigure{\includegraphics[width=0.49\textwidth]{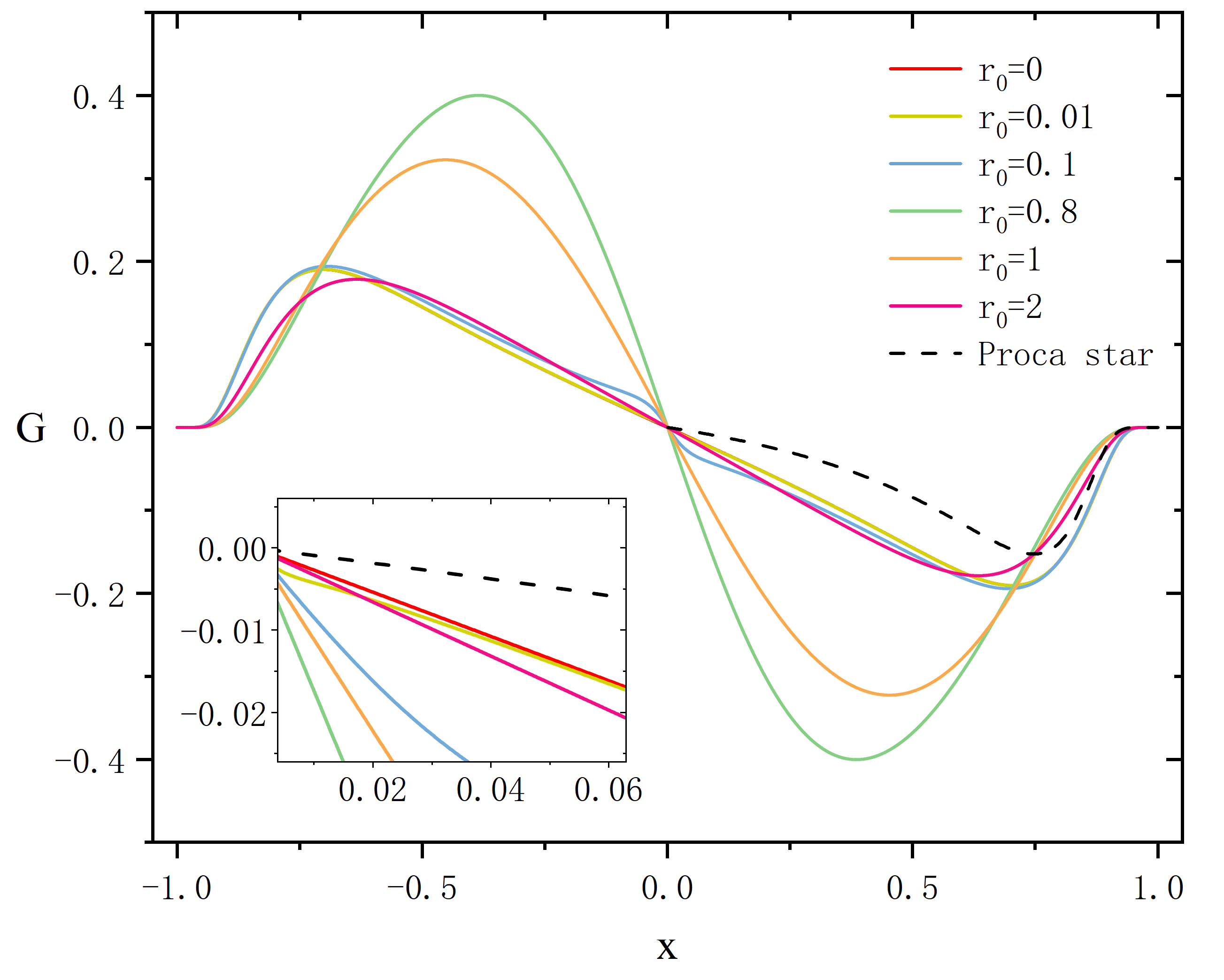}}
  \end{center}
\caption{In the case of the symmetric type I, the Proca field $F$ (left) and $G$ (right) as functions of $x$ with different values of $r_{0}$ for $\omega=0.85$. The dashed line corresponds to the field functions of an independent Proca Star. Since $r_{0}$ is very small, the function curves are very close to each other, so we use subfigures to show the function curve distribution near $x=0$.}
\label{fg1}
\end{figure}

For the first type of the symmetric solution, the numerical results of the Proca field function at $\omega=0.85$ are shown in Fig. \ref{fg1}. It can be seen that the field function $F$ has two nodes and $G$ has no nodes (excluding those where the field function value is zero at the origin). When $r_{0}$ takes different values, all solutions are in the first branch or there is only one branch. The field function $F$ is symmetric about $x=0$ and reaches the maximum value at $x=0$. As $r_{0}$ increases, the maximum value of $F$ first increases and then decreases. The field function $G$ passes through the origin and is antisymmetric about the origin and as $r_{0}$ increases, the slope at $x=0$ first increases and then decreases.

\begin{figure}
  \begin{center}
\subfigure{\includegraphics[width=0.49\textwidth]{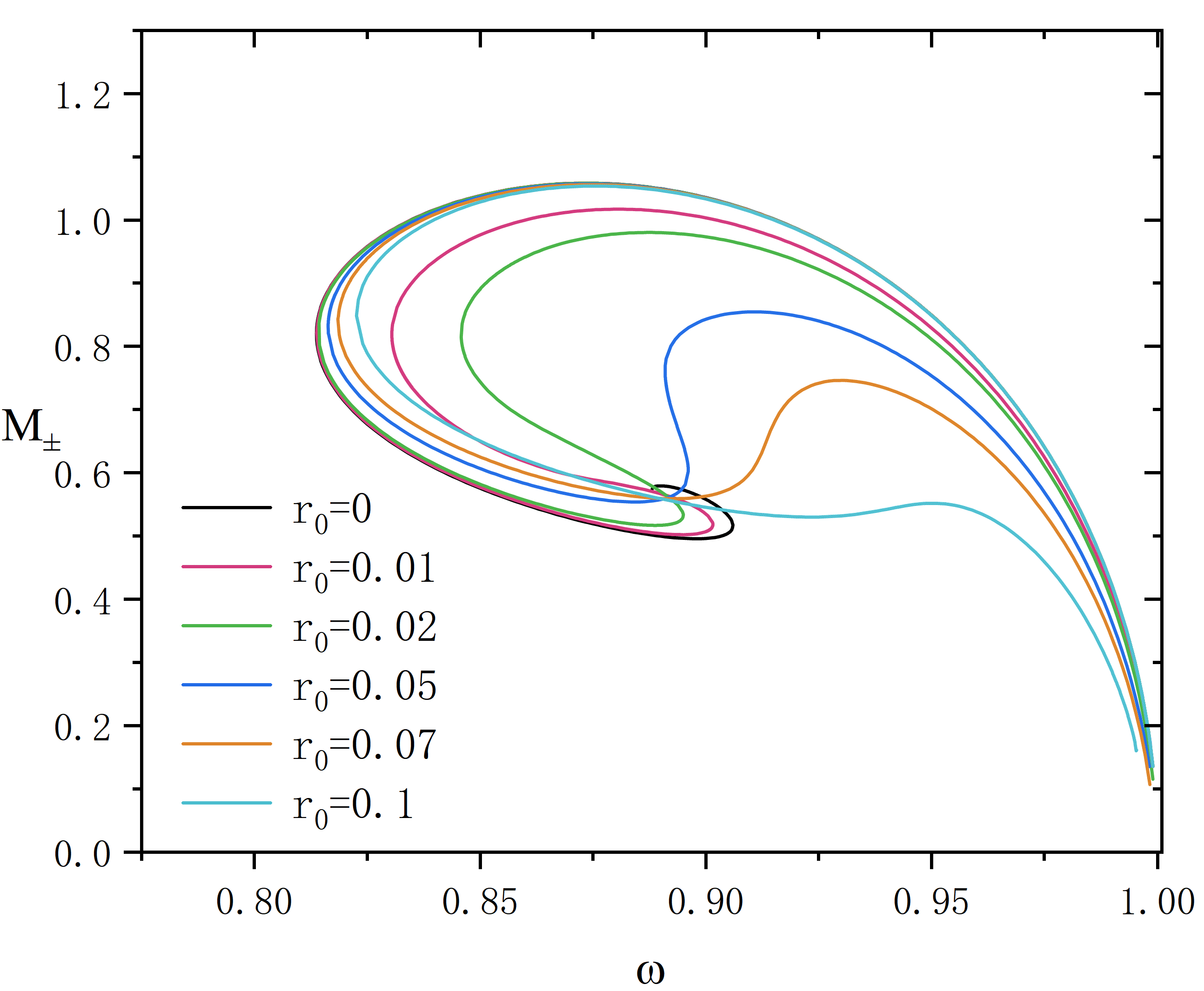}}
\subfigure{\includegraphics[width=0.49\textwidth]{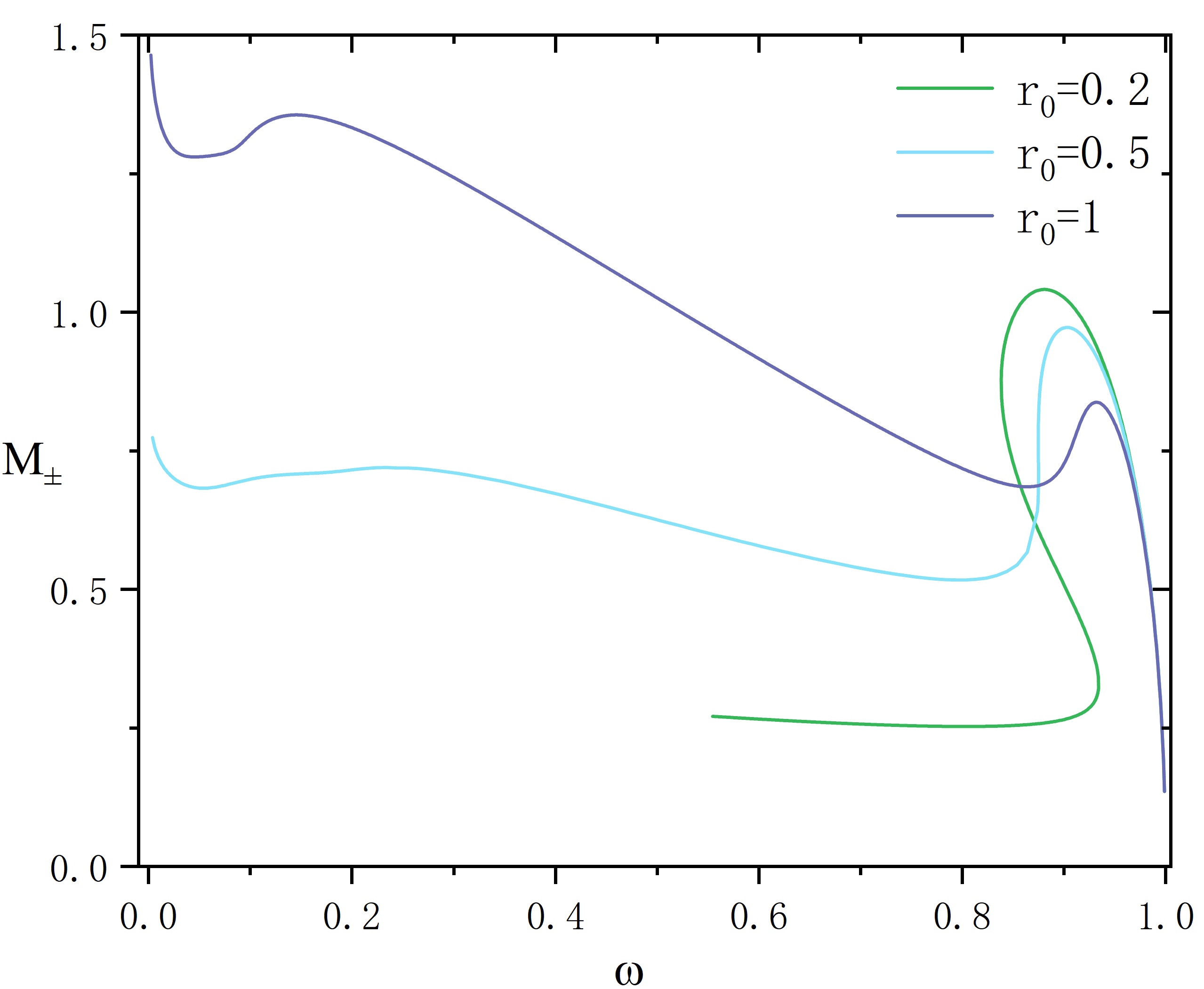}}
\subfigure{\includegraphics[width=0.49\textwidth]{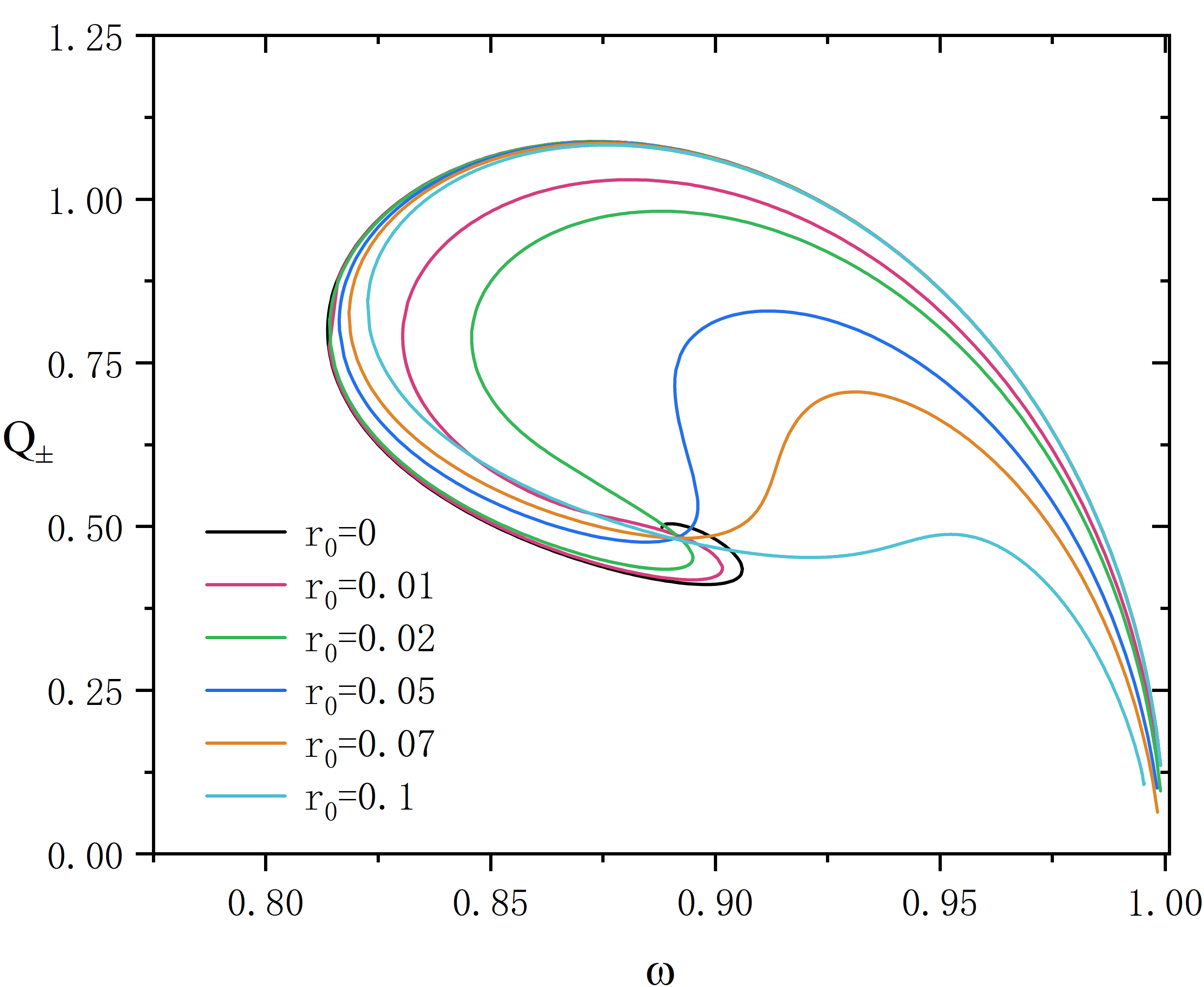}}
\subfigure{\includegraphics[width=0.49\textwidth]{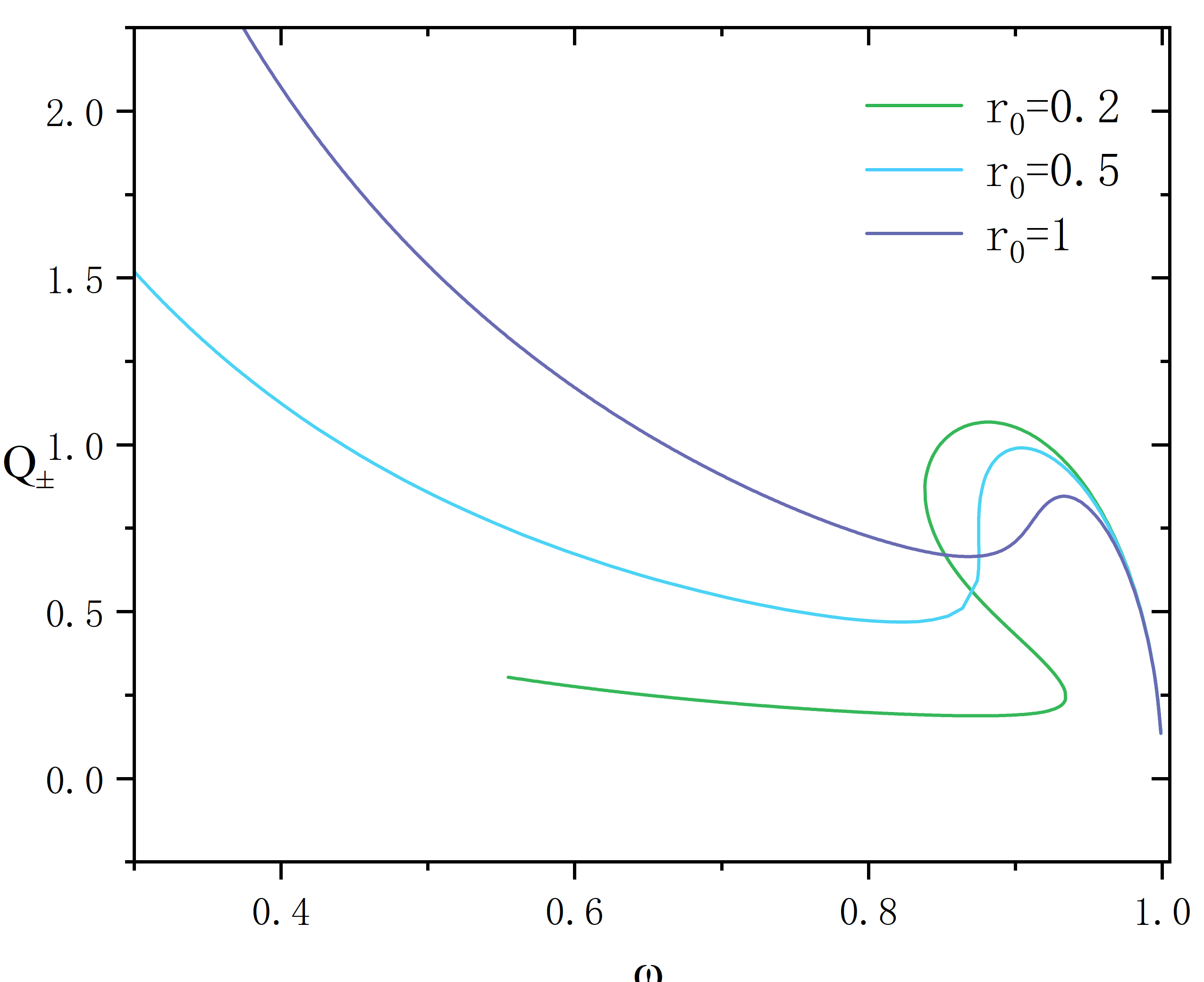}}
  \end{center}
\caption{In the case of the symmetric type I, the mass $M$ and the charge $Q$ as functions of frequency $\omega$ for different values of $r_{0}$ .}
\label{mq1}
\end{figure}

Fig. \ref{mq1} shows the images of mass $M$ and charge $Q$ of frequency $\omega$ when $r_{0}$ takes different values. Since the solution has symmetry, $M_{+} = M_{-}$, $Q_{+} = Q_{-}$. The curve of $r_{0}=0$ corresponds to the case without the phantom field. $M$ and $Q$ have the same magnitude and their curve shapes are similar to each other.

In particular, when $r_{0}$ is small, the results are distributed in the narrow range of $\omega$, the solution will become a multi-branch solution, and the curve shape will show a loop structure. As the value of $r_{0}$ increases, the loop structure of the solution disappears, and the high branch and the low branch gradually combine together. The final result is only one branch, such as the result of $r_{0}=0.5,1$ and the domain of functions is expanded. The larger $r_{0}$ is, the smaller the minimum value of $\omega$ can reach. Because the value of $Q$ is very large when $\omega$ is close to zero, we do not show all the results in the figure. Interestingly, when $r_{0}=0$, the curves of $M$ and $Q$ coincide with the solution of a Proca star, and the field functions are consistent with the field functions of a Proca star in the region $x>0$. Since $F$ and $G$ are symmetric and antisymmetric about $x=0$, it can be understood that at infinity, the system degenerates into two individual Proca stars.

\begin{figure}
  \begin{center}
\subfigure{\includegraphics[width=0.5\textwidth]{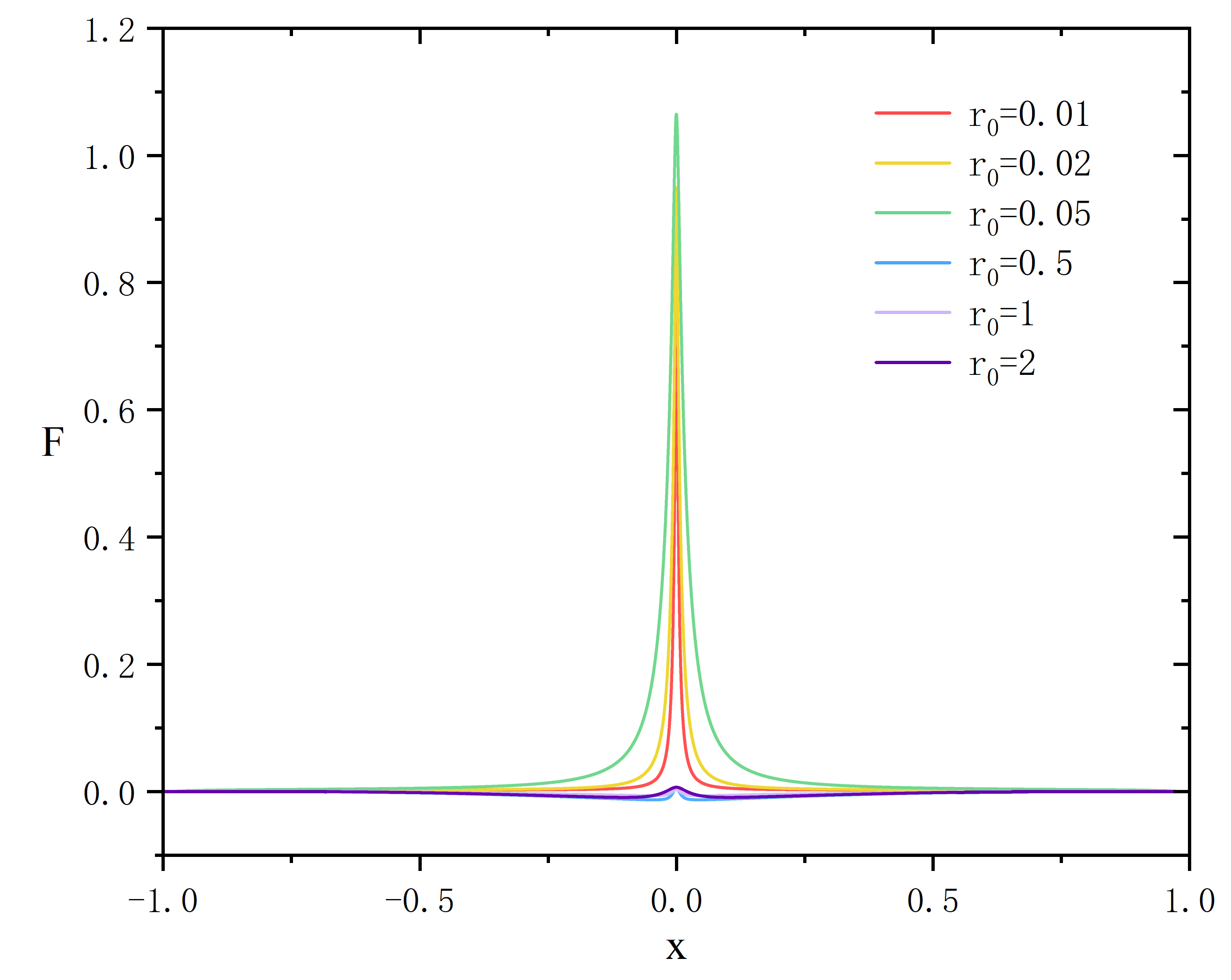}}
\subfigure{\includegraphics[width=0.48\textwidth]{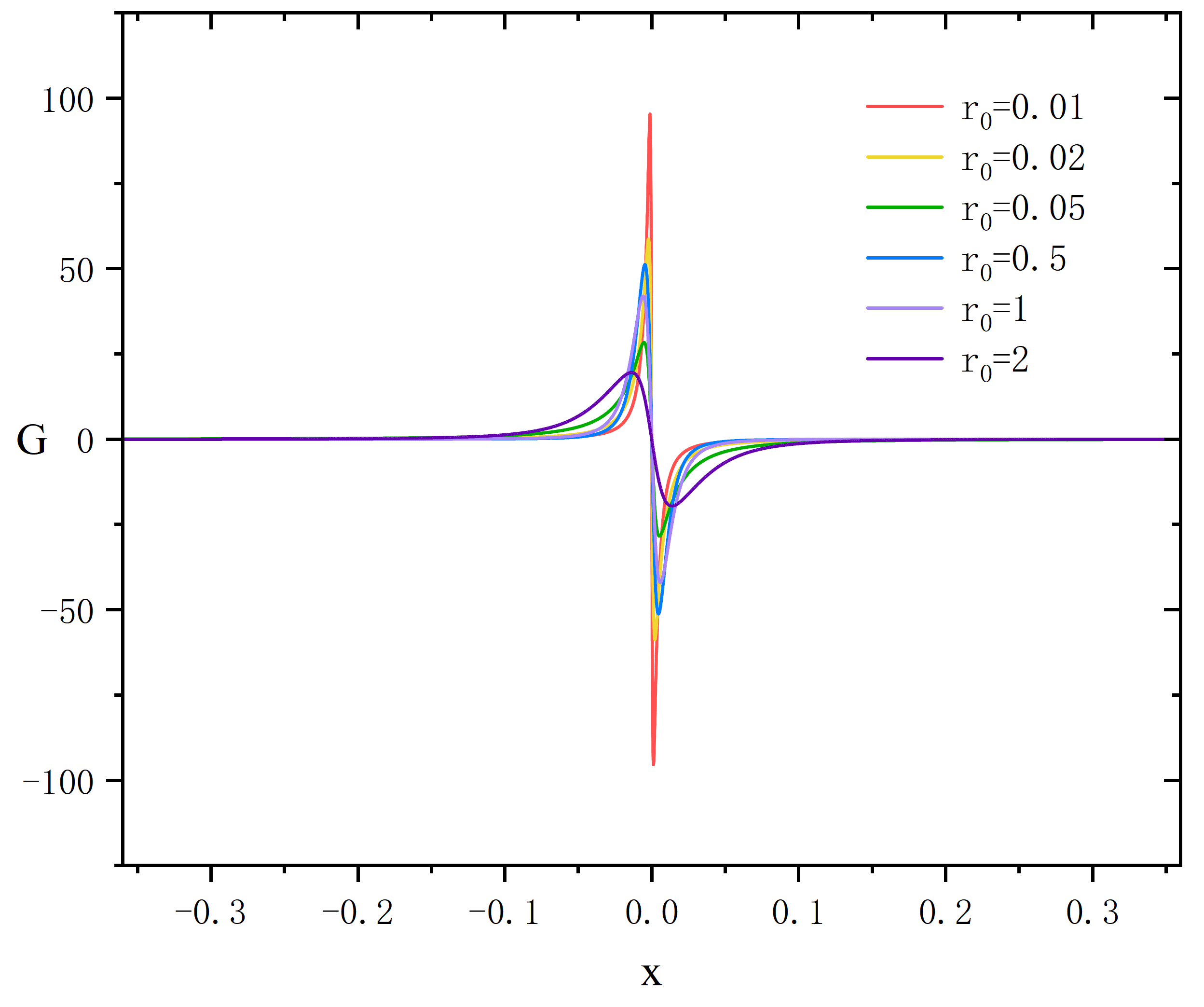}}
  \end{center}
\caption{In the case of the symmetric type I, the Proca field $F$ (left) and $G$ (right) as functions of $x$ when the frequency $\omega$ takes the endpoint of the last branch.}
\label{fg11}
\end{figure}

\begin{figure}
  \begin{center}
\subfigure{\includegraphics[width=0.49\textwidth]{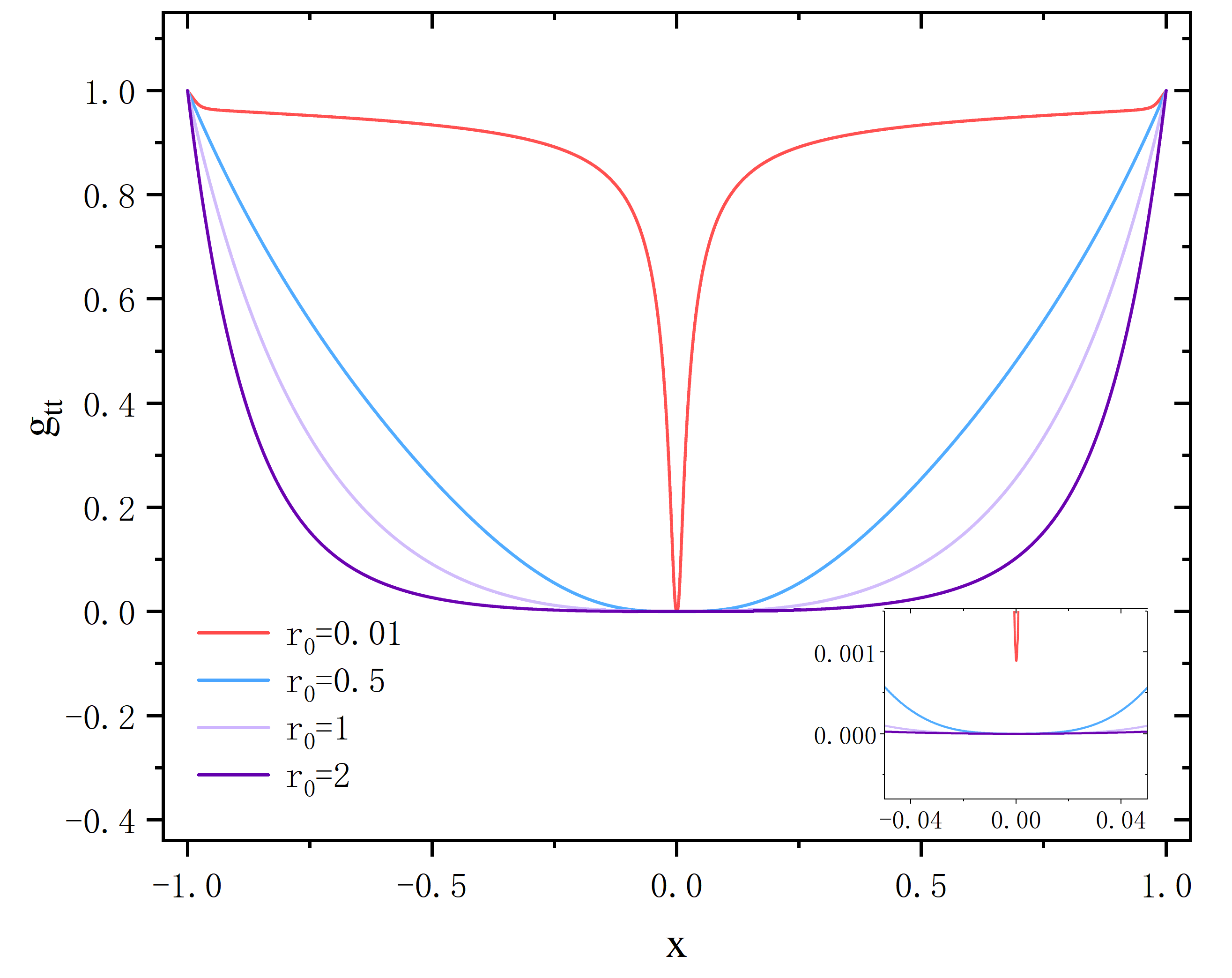}}
\subfigure{\includegraphics[width=0.49\textwidth]{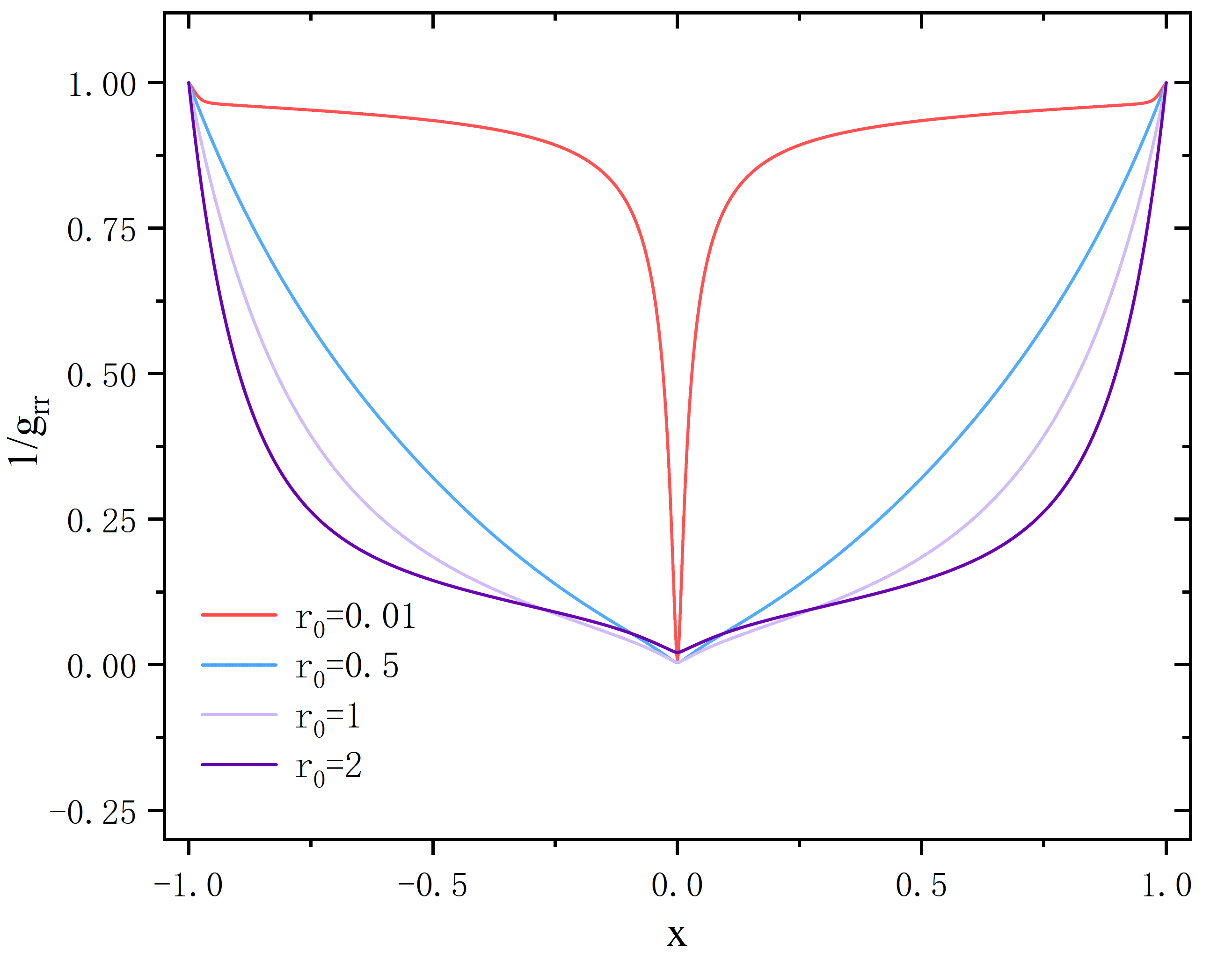}}
  \end{center}
\caption{$g_{tt}$ and $1/g_{rr}$ as functions of $x$ for the symmetric type I.}
\label{gtr1}
\end{figure}

    	\begin{table}[ht] 
	\centering 
     \setlength{\tabcolsep}{16mm}
	\begin{tabular}{|c||c|c|}
\hline
		$r_0$ & $\omega$ & $g_{tt}(min)$ \\
\hline
		$0.01$ & $0.9999$ & $9.0\times10^{-4}$ \\
\hline
		$0.5$ & $0.02$ & $2.3\times10^{-6}$ \\
\hline
		$1$ & $0.01$ & $8.4\times10^{-7}$ \\
\hline
		$2$ & $0.01$ & $2.4\times10^{-6}$ \\
\hline
	\end{tabular}
 	\caption{Under different values of $r_0$, the minimum mertic $g_{tt}$ in the specific frequency.}
	\label{tab:t1}
\end{table}

\begin{figure}
  \begin{center}
\subfigure{\includegraphics[width=0.49\textwidth]{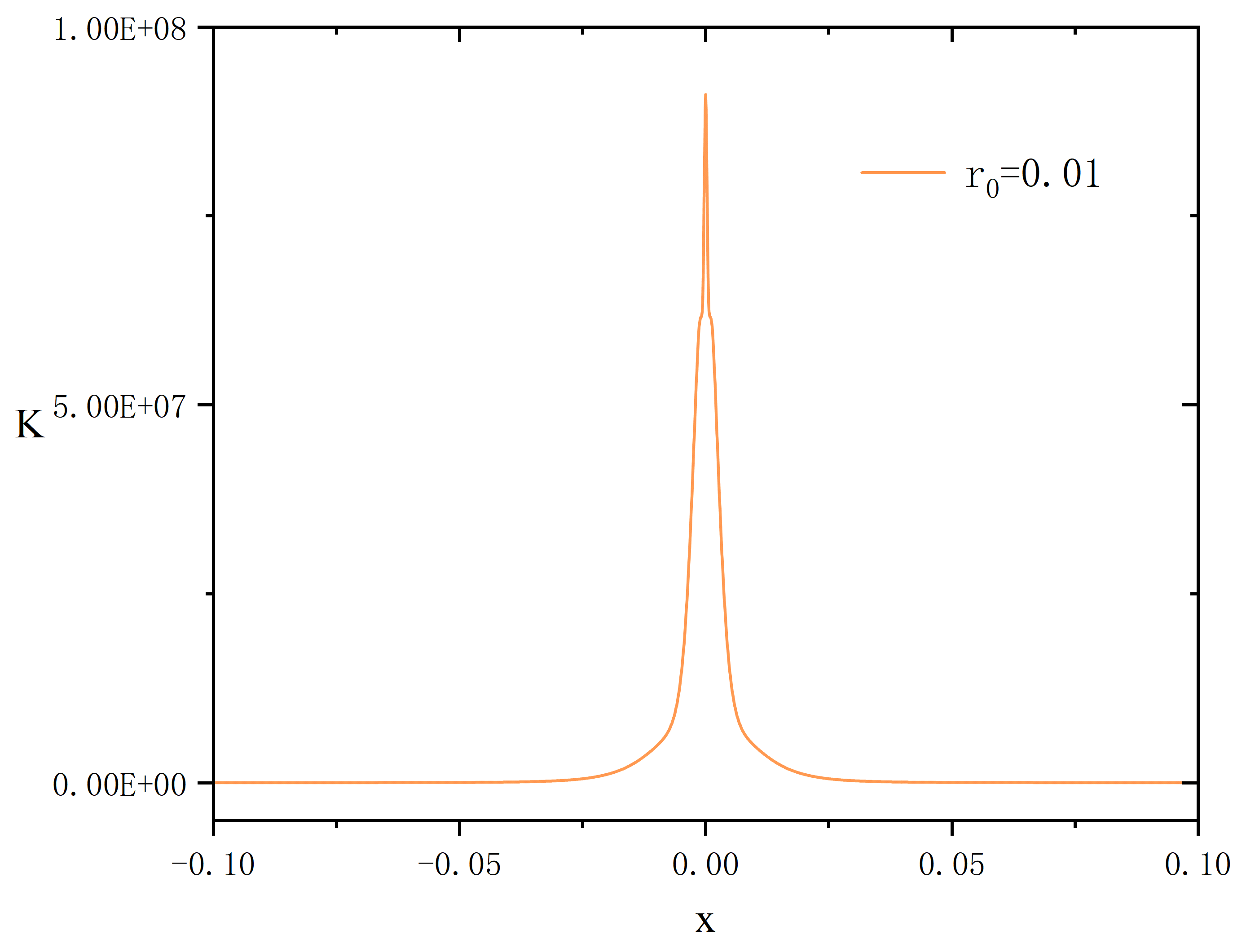}}
\subfigure{\includegraphics[width=0.48\textwidth]{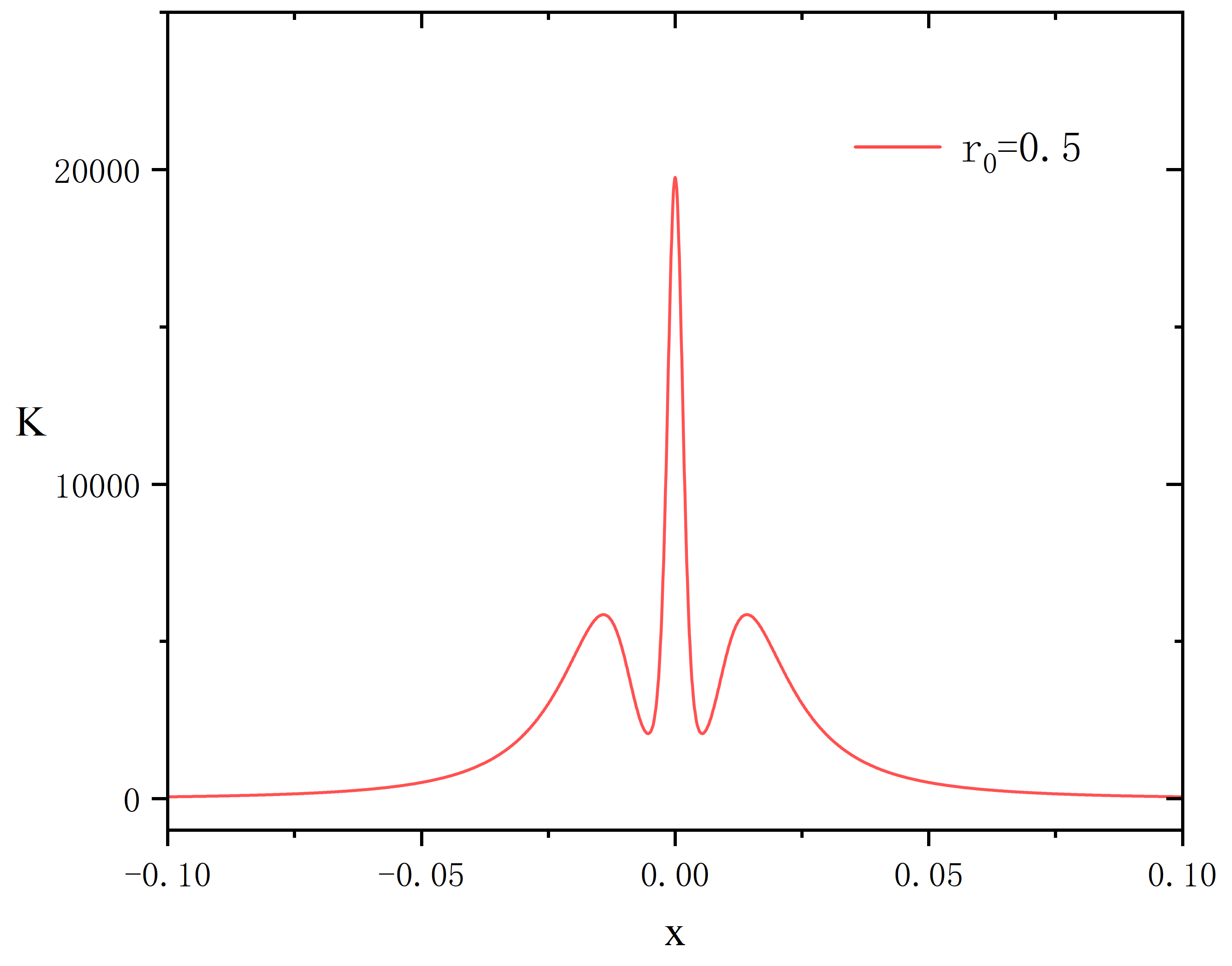}}
\subfigure{\includegraphics[width=0.49\textwidth]{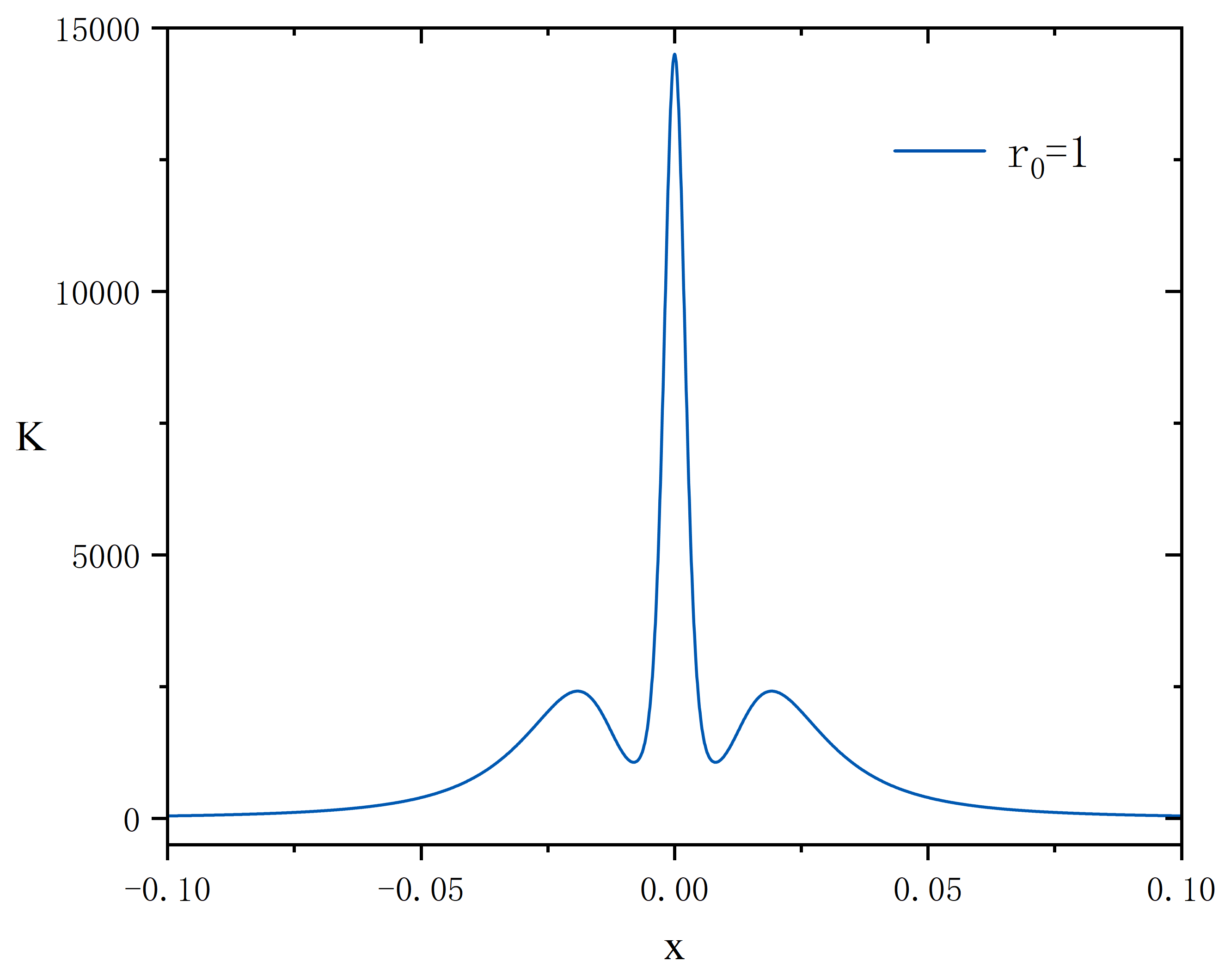}}
\subfigure{\includegraphics[width=0.49\textwidth]{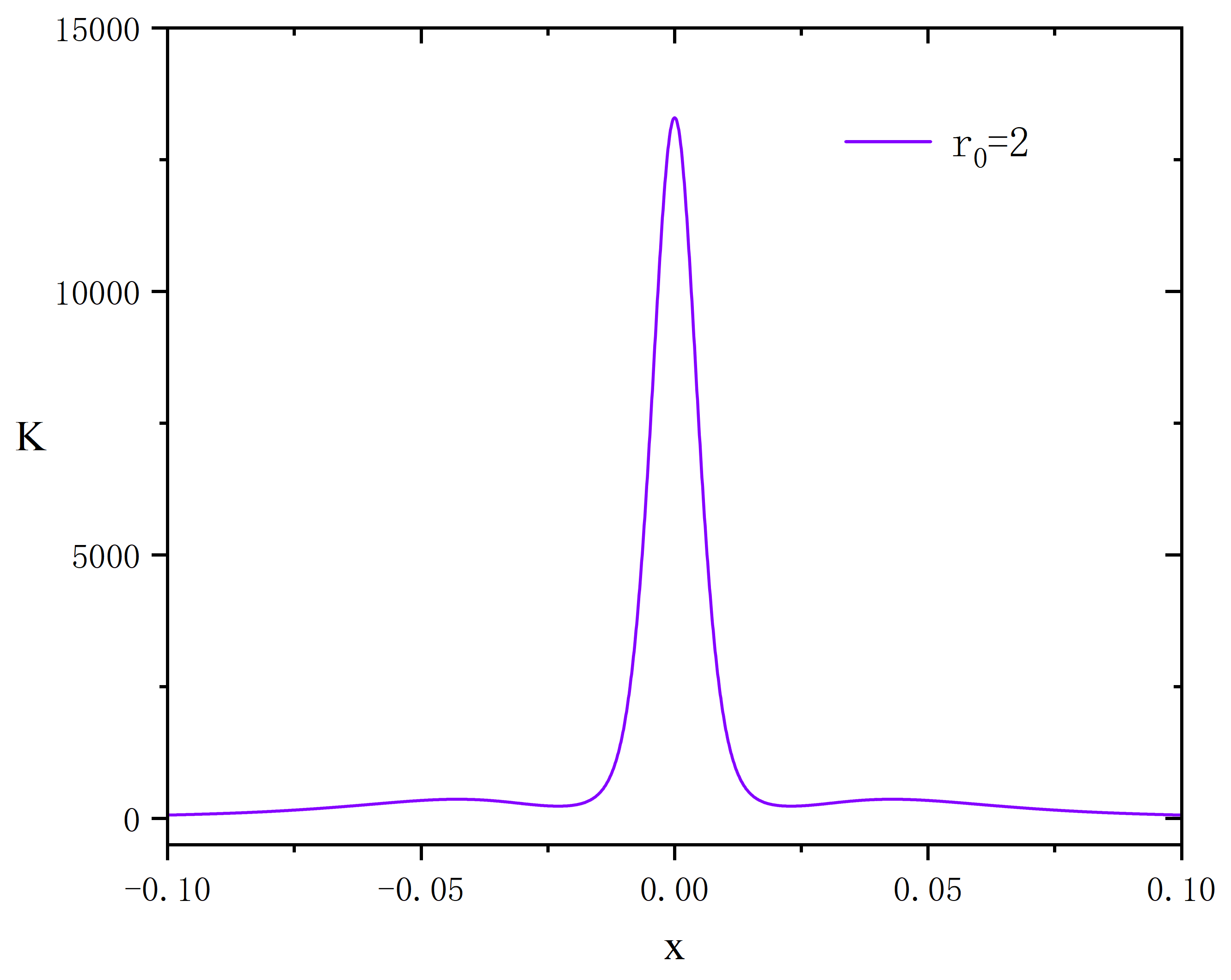}}
  \end{center}
\caption{The distribution of the Kretschmann scalar for the symmetric type I.}
\label{k1}
\end{figure}

Furthermore, we found that the wormhole system has an extremely approximate black hole solution at $x=0$. This phenomenon happens near the end of the last branches, and as the frequency $\omega$ approaches the end point, the appearance of an extremely approximate black hole becomes increasingly apparent. Moreover, Fig. \ref{fg11} shows that the field function is concentrated near $x=0$ at the end of the last branch, and the smaller the $r_{0}$, the sharper the field function. Then we analyze the metric function $g_{tt}$ and $g_{rr}$ as shown in Fig. \ref{gtr1}, at $x=0$ the value of the metric function $g_{tt}$ will reach the minimum value and tend to zero. At the same time, the value of $g_{rr}$ tends to infinity, i.e., the reciprocal of $g_{rr}$ tends to zero. Tab. \ref{tab:t1} shows the minimum value of $g_{tt}$ corresponding to different $r_{0}$ and the frequency in detail. It can be found that the minimum value of $g_{tt}$ is all less than $10^{-3}$, and some can even be less than $10^{-6}$. We consider an event horizon appears when the value of $g_{tt}$ is less than $10^{-3}$, which is a sign of the emergence of extreme approximate black holes. Therefore, for all values of $r_{0}$, an extremely approximate black hole solution will appear.

In article \cite{Morris:1988cz}, the work of M. Morris and K. Thorne put some restrictions on traversable wormholes, such as no horizon and no singularity. An important symbol reflecting the singularity is the Ricci tensor or the Kretschmann scalar\cite{Dymnikova:1992ux,Ayon-Beato:1998hmi}. Because a horizon appears in our wormhole model, we choose to calculate the Kretschmann scalar to analyze the singularity of the system. As shown in Fig. \ref{k1}, we selected four groups of $r_{0}$ and corresponding $\omega$ to calculate the Kretschmann scalar and found that the Kretschmann scalar diverges when $x=0$, and as $r_{0}$ increases, the degree of divergence decreases. This can show that the singularity occurs at $x=0$ and also indicates that the solution corresponds to an untraversable wormhole. We consider the position where the Kretschmann scalar diverges as the position of the event horizon. Therefore, it can be concluded that in the symmetric type \uppercase\expandafter{\romannumeral1}, an extremely approximate black hole will appear at $x=0$. At this time, the matter field is concentrated near the event horizon.

\subsection{Symmetric type \uppercase\expandafter{\romannumeral2} }

For the second type of symmetry solutions, the numerical results of the Proca field function at $\omega=0.85$ are shown in Fig. \ref{fg2}. The functions of $F$ and $G$ both have two nodes, excluding those where the field function value is zero at the origin. It can be seen that at the position where $x$ is larger, the curves for different $r_{0}$ have little difference and basically overlap. In the area where $x$ is smaller, the shape of the curves differs much when $r_{0}$ is different. The field function $F$ is antisymmetric about $x=0$, the maximum value of the function increases with the decrease of $r_{0}$, and the extreme point converges to the origin $x=0$. The field function $G$ is symmetric about $x=0$ and reaches the maximum value at $x=0$, as  $r_{0}$ decreases, the maximum of $G$ increases.

\begin{figure}
  \begin{center}
\subfigure{\includegraphics[width=0.49\textwidth]{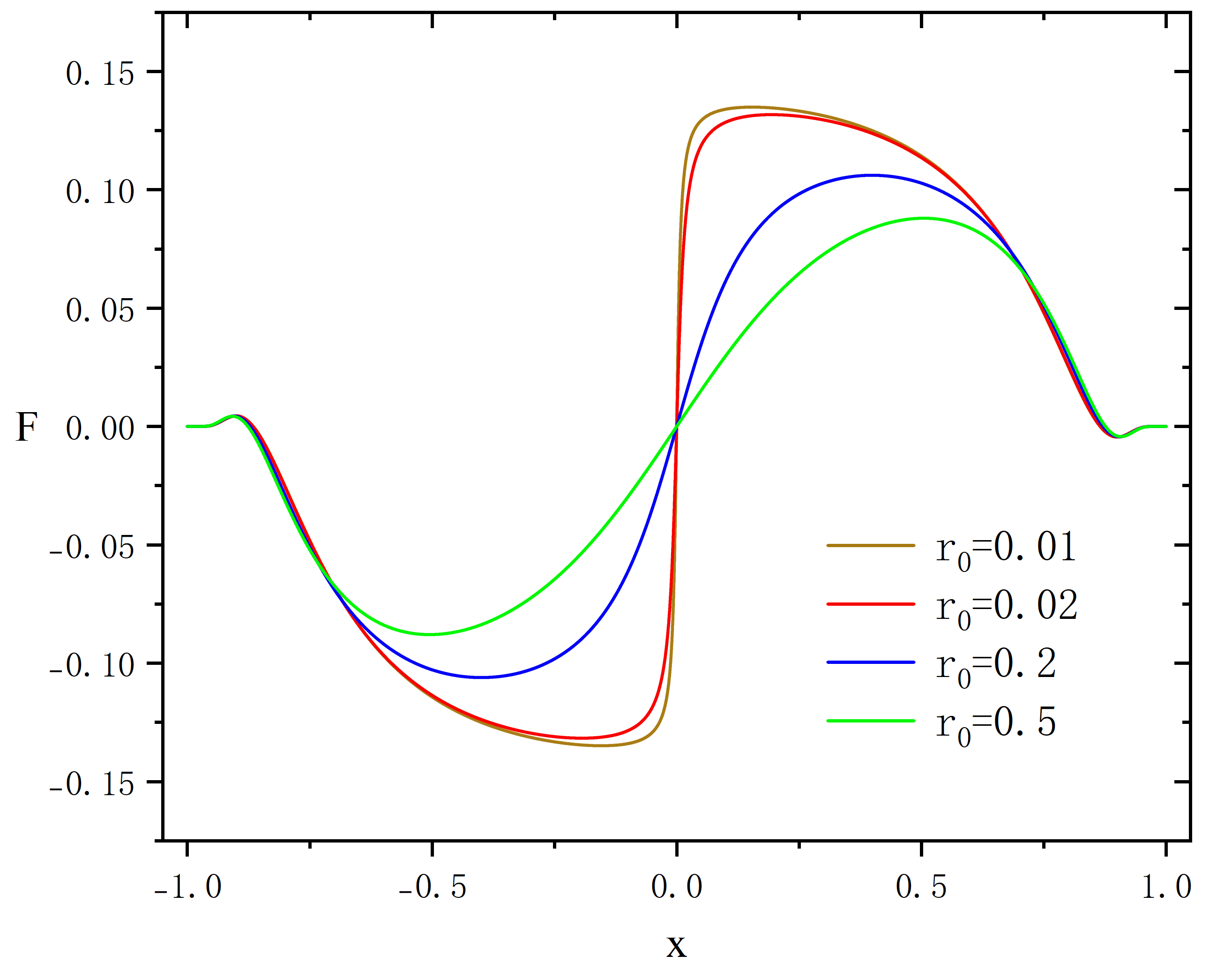}}
\subfigure{\includegraphics[width=0.49\textwidth]{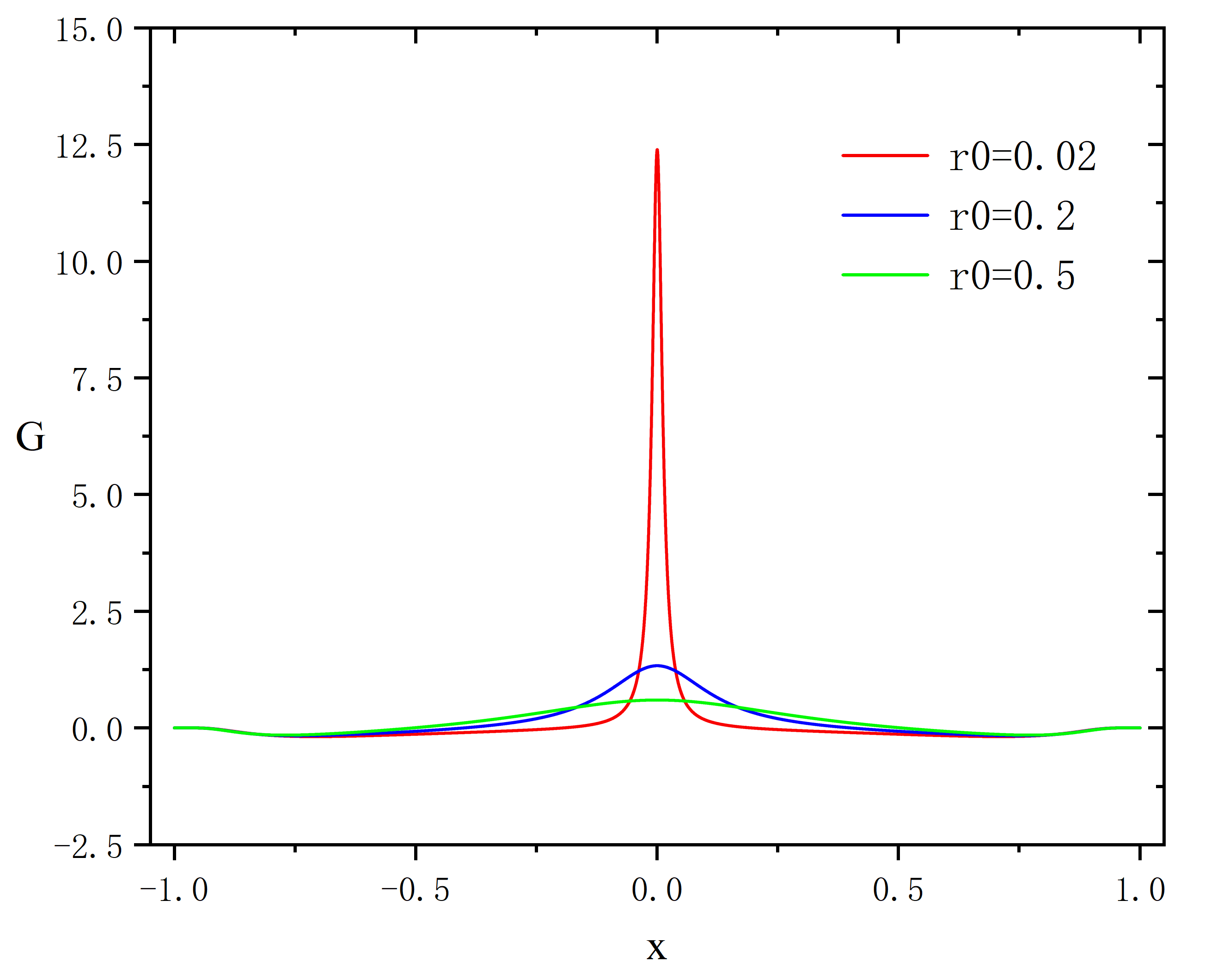}}
  \end{center}
\caption{In the case of the symmetric type II, the Proca field $F$ (left) and $G$ (right) as functions of $x$ with different values of $r_{0}$ for $\omega$ = 0.85.}
\label{fg2}
\end{figure}

In Fig. \ref{mq2}, we show the ADM mass $M$ and the Noether charge $Q$ as a function of the frequency $\omega$ when the throat size $r_{0}$ takes different values. The black curve corresponds to the Proca star solution. When $r_{0}$ increases from 0, the curve will first form a spiral structure and then a loop. As $r_{0}$ continues to increase, the loop characteristics of the curve disappear and the multi-valued curve will become a single-valued curve. In the first branch, when the frequency is fixed, we find that as $r_{0}$ increases, the values of both mass $M$ and charge $Q$ increase, and they have similar behavior. The solution in this case cannot degenerate into a single Proca star. 

\begin{figure}
  \begin{center}
\subfigure{\includegraphics[width=0.49\textwidth]{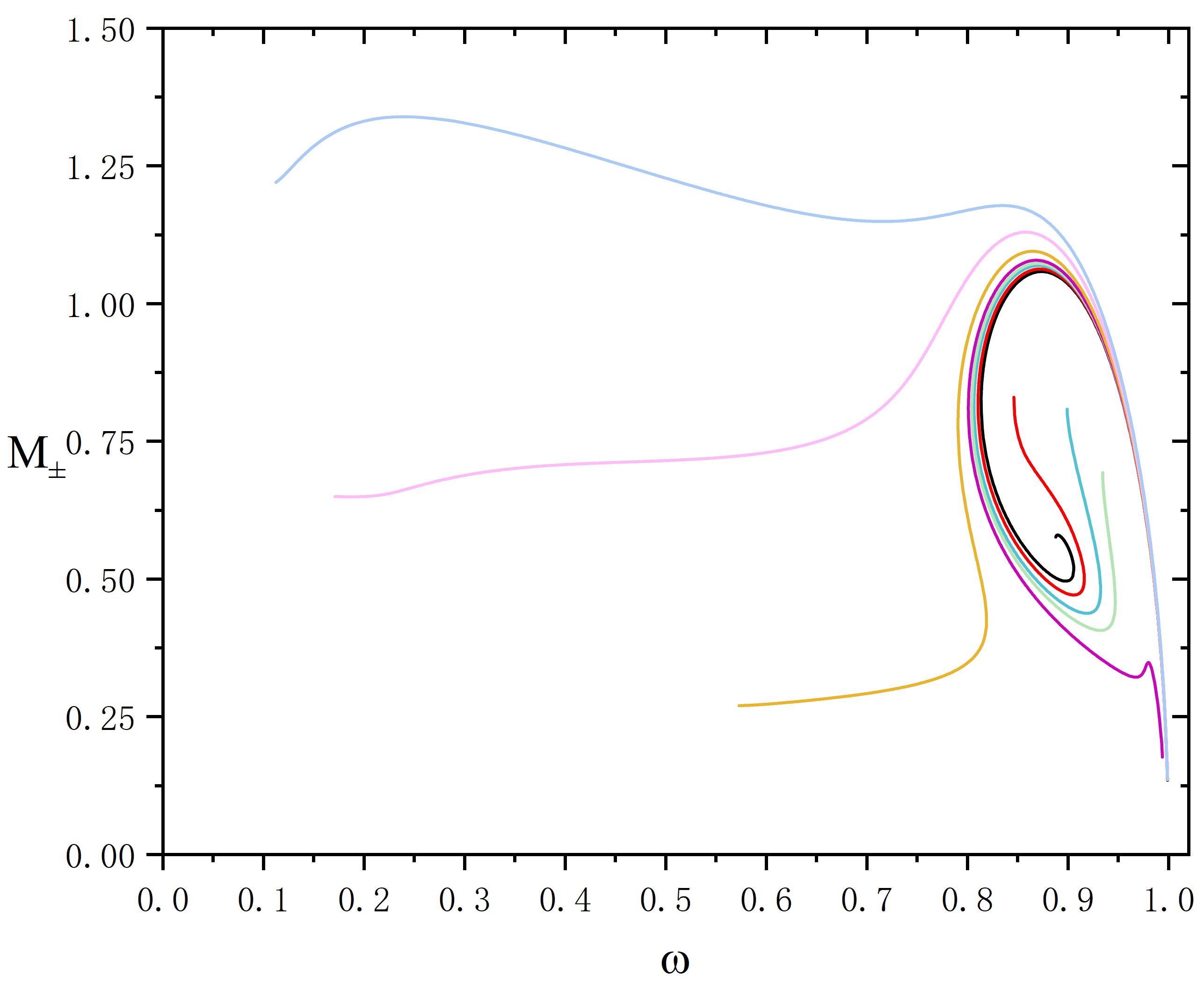}}
\subfigure{\includegraphics[width=0.49\textwidth]{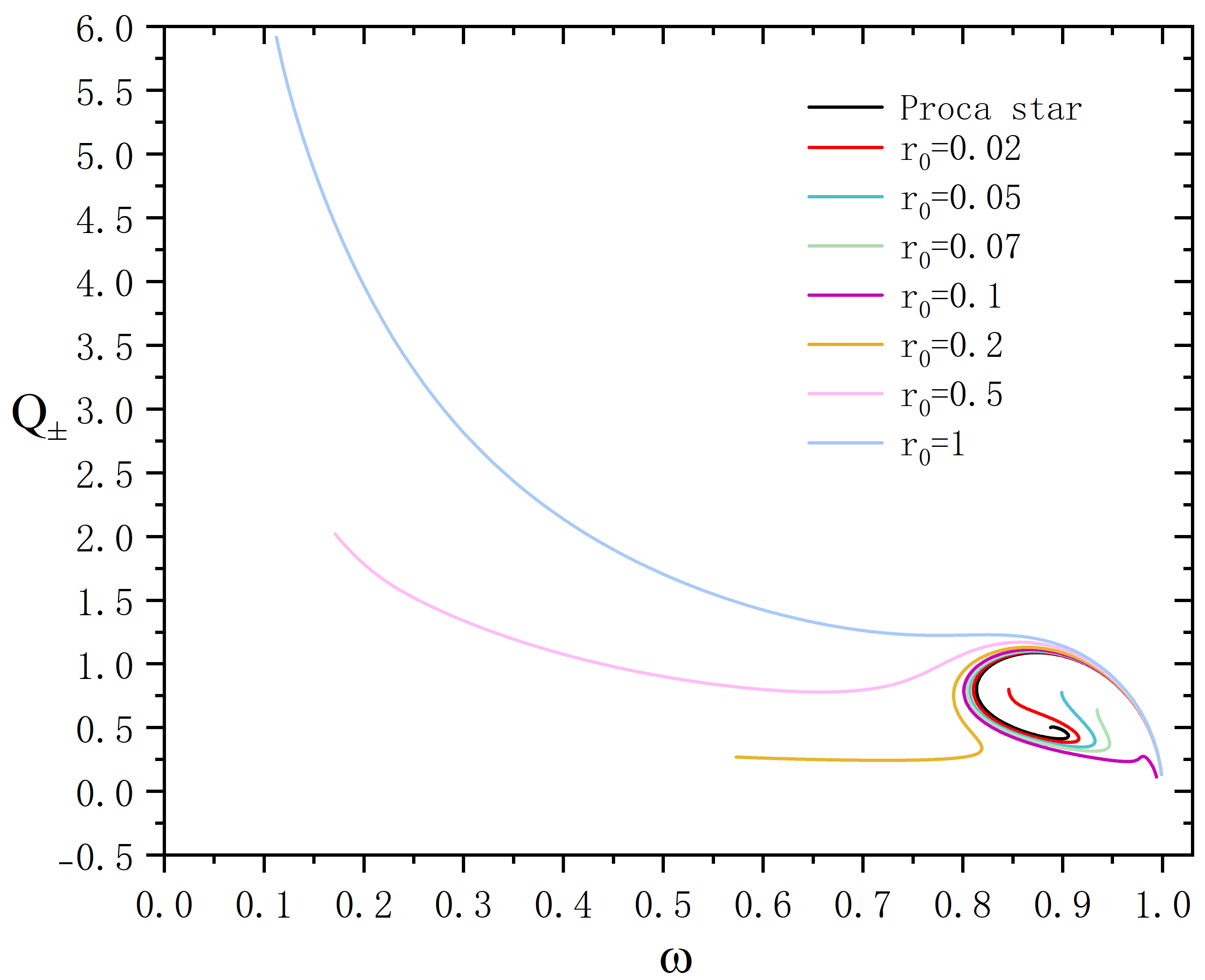}}
  \end{center}
\caption{In the case of the symmetric type II, the mass $M$ and the charge $Q$ as functions of frequency $\omega$ for different values of $r_{0}$. The parameters of each curve in the left picture are consistent with the right picture.}
\label{mq2}
\end{figure}

\begin{figure}
  \begin{center}
\subfigure{\includegraphics[width=0.48\textwidth]{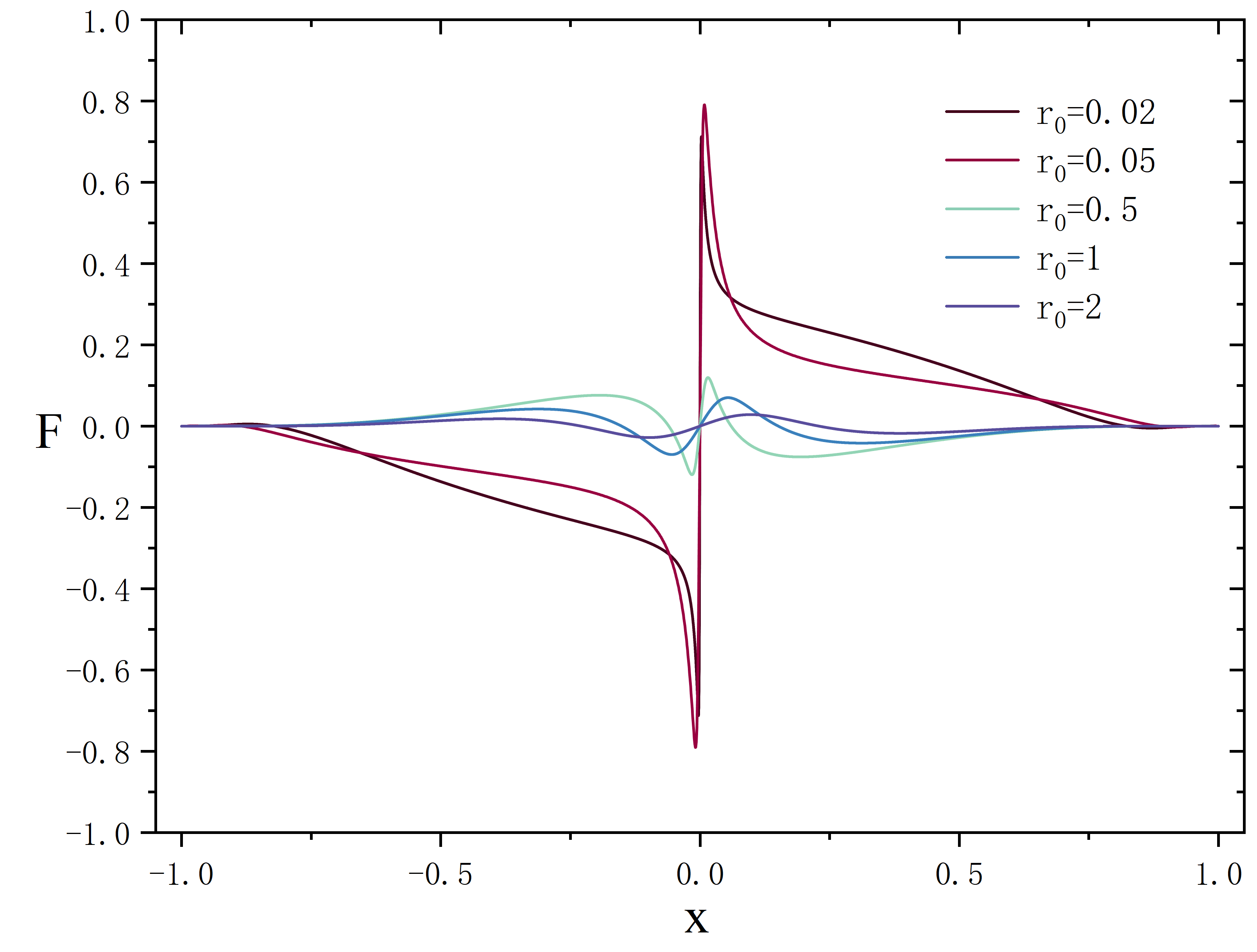}}
\subfigure{\includegraphics[width=0.48\textwidth]{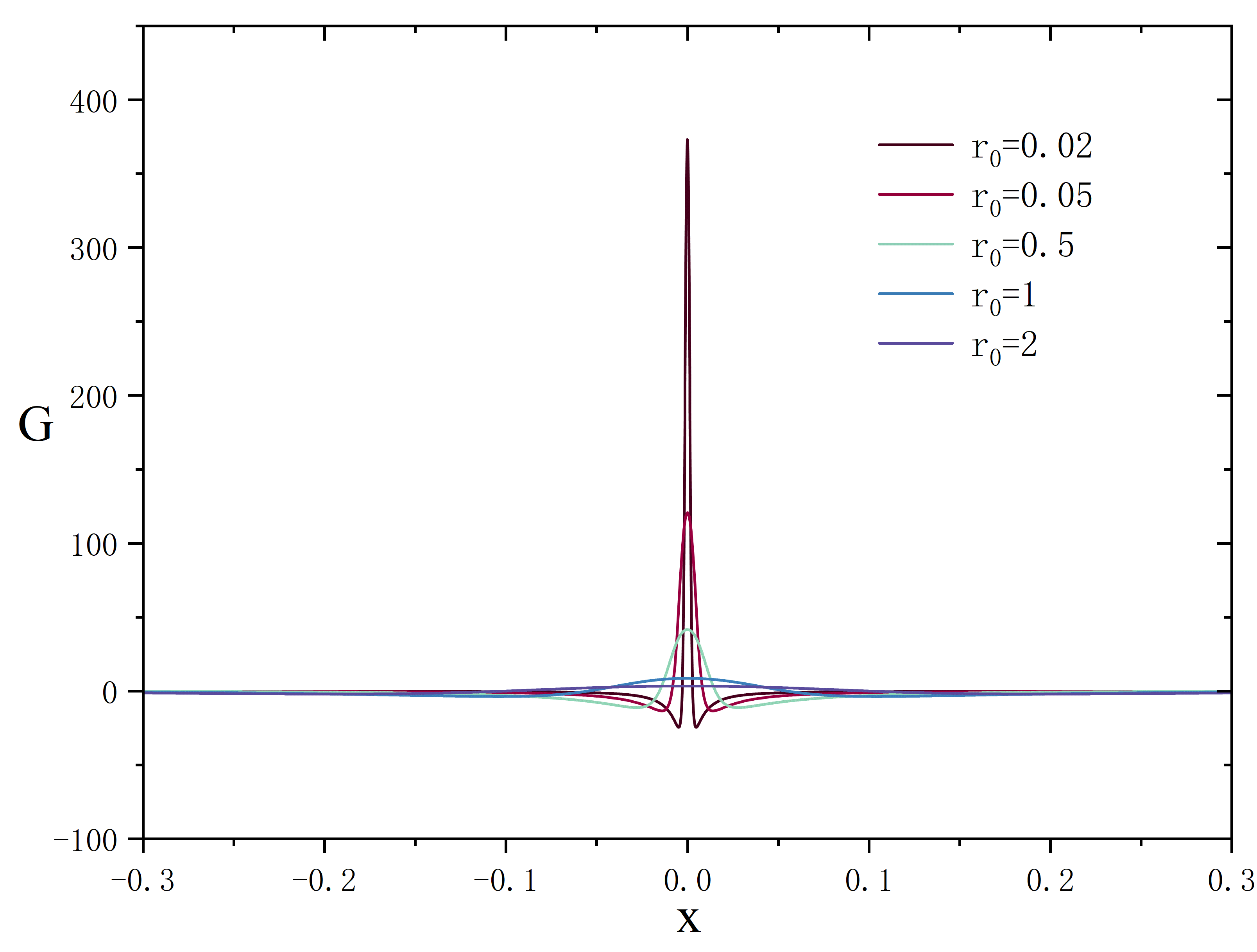}}
  \end{center}
\caption{In the case of the symmetric type II, the Proca field $F$ (left) and $G$ (right) as functions of $x$ when the frequency $\omega$ takes the endpoint of the last branch.}
\label{fg22}
\end{figure}

\begin{figure}
  \begin{center}
\subfigure{\includegraphics[width=0.49\textwidth]{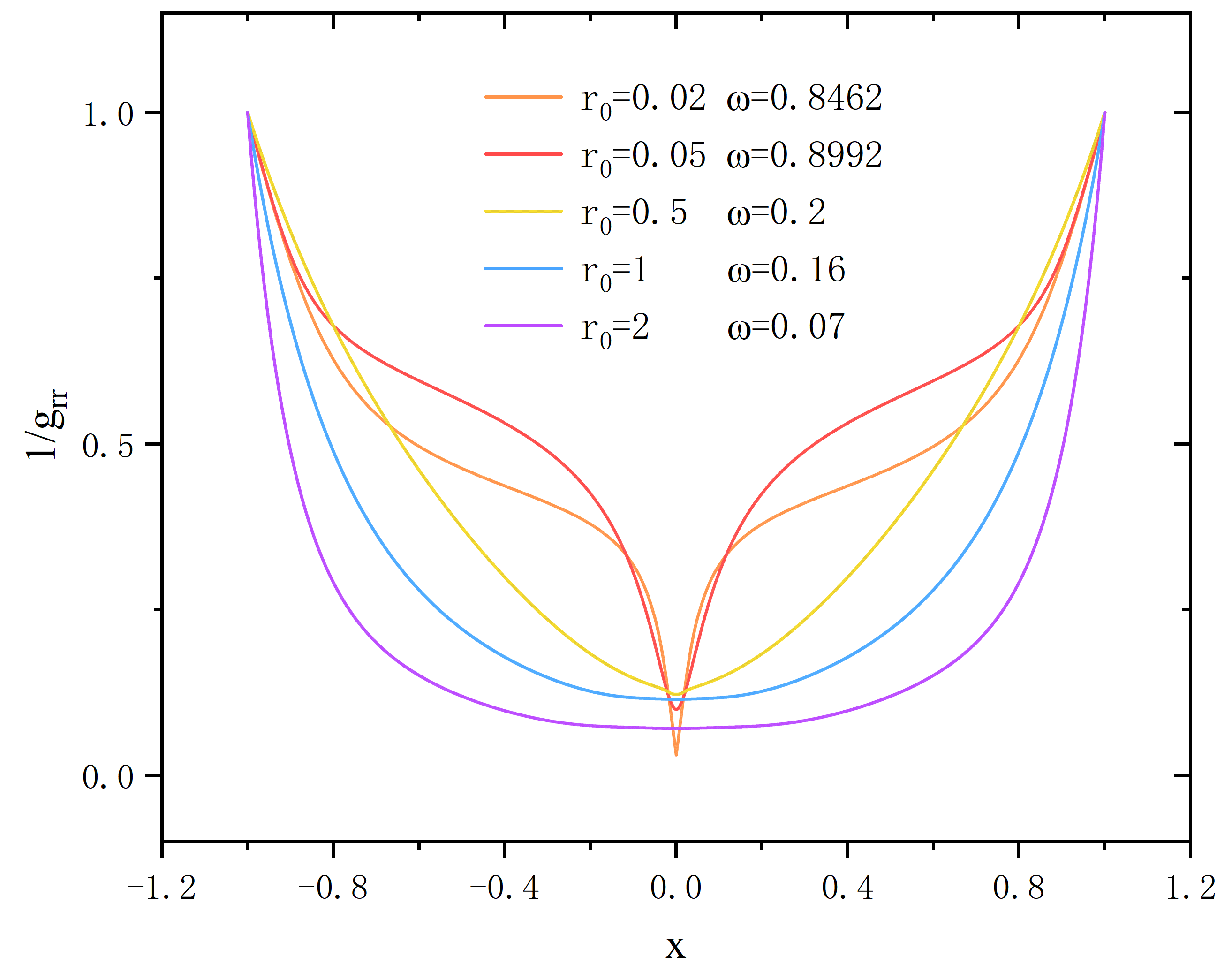}}
\subfigure{\includegraphics[width=0.49\textwidth]{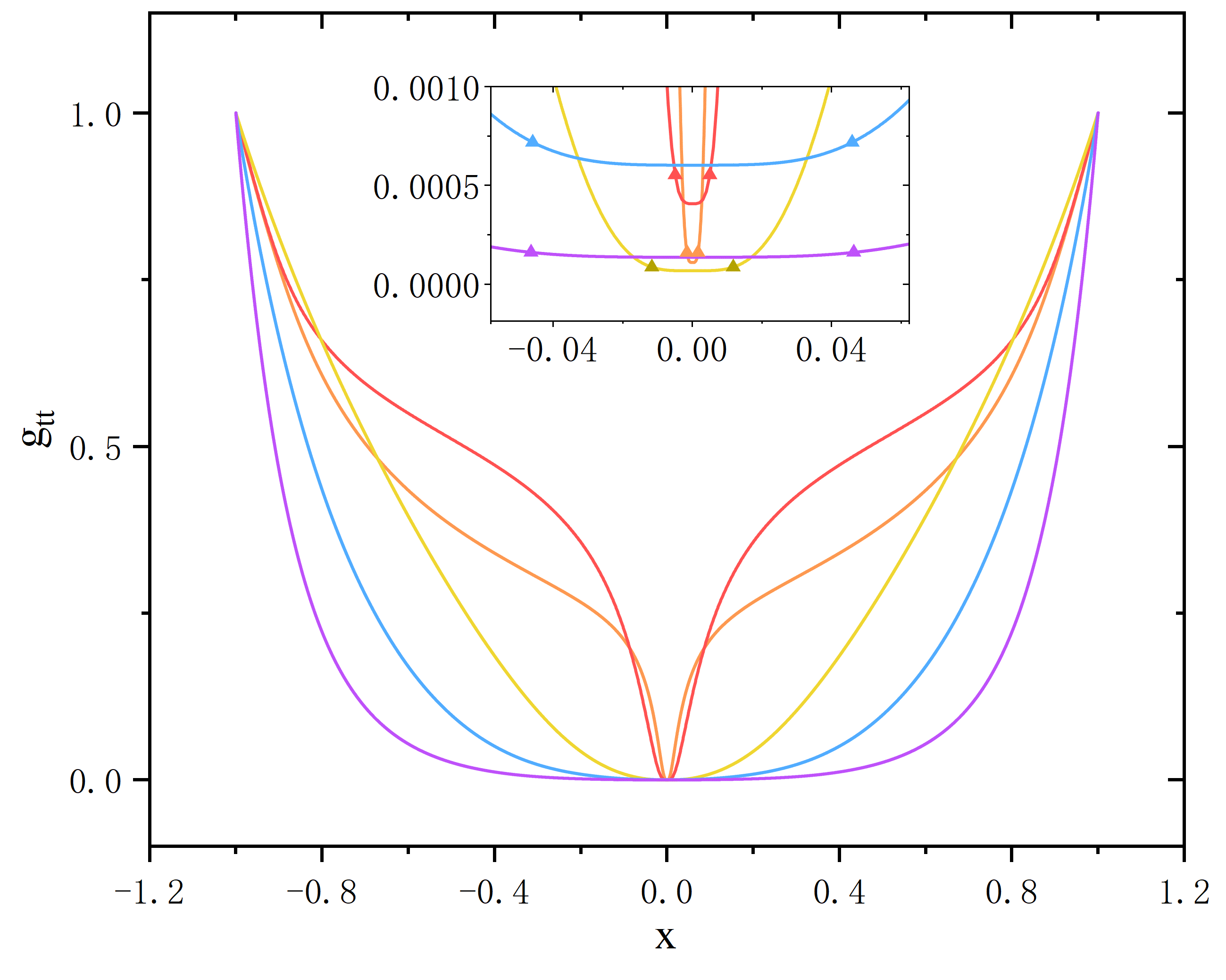}}
  \end{center}
\caption{$g_{tt}$ and $1/g_{rr}$ as functions of $x$ for the symmetric type II. The triangle symbol represents the location of the event horizon. The parameters of each curve in the right picture are consistent with the left picture.}
\label{gtr2}
\end{figure}

We reasonably suspect that in this symmetry case, there are also extremely approximate black hole solutions. We show functions of the Proca field when the frequency $\omega$ takes the endpoint of the last branch in Fig. \ref{fg22}. From the graphs, we can see that the field distribution is concentrated near $x=0$, and the smaller $r_{0}$ is, the more concentrated the field distribution is. It can be seen that in Fig. \ref{gtr2} the metric function $g_{tt}$ and $1/g_{rr}$ both tend to zero at the specific frequency. Different from the symmetric type I, the metric function $g_{tt}$ will reach the same minimum value in a region and as the throat size increases, the region becomes larger. Under different parameters, the minimum value of $g_{tt}$ and its distribution range are shown in the Tab. \ref{tab:t2}. It can be found that almost all minimum values of $g_{tt}$ reach $10^{-4}$. So we think that no matter what value $r_{0}$ takes, there will always be a black hole solution at a specific frequency. Next, we calculate the Kretschmann scalar, as shown in Fig. \ref{k2}. We find that the Kretschmann curve has two peaks symmetrical about $x=0$. The area between the two peaks is close to the $x$ distribution area corresponding to the minimum value of $g_{tt}$. We stipulate that the $x$ coordinate corresponding to the peak of the Kretschmann curve is the position of the event horizon, then no matter what the size of the wormhole throat is, an extremely approximate black hole will appear symmetrically on both sides of the throat. Considering the above reasons, we call this solution a black hole-wormhole-black hole combination. At this time, the wormhole is untraversable and the matter field is concentrated near the event horizon.

 	\begin{table}[ht] 
	\centering 
     \setlength{\tabcolsep}{16mm}
	\begin{tabular}{|c||c|c|}
\hline
		$r_0$ & $x$ & $g_{tt}(min)$ \\
\hline
		$0.02$ & $-0.001\le x\le 0.001$ & $1.1\times10^{-4}$ \\
\hline
		$0.05$ & $-0.002\le x\le 0.002$ & $4.1\times10^{-4}$ \\
\hline
		$0.5$ & $-0.003\le x\le 0.003$ & $6.7\times10^{-5}$ \\
\hline
		$1$ & $-0.016\le x\le 0.016$ & $6.0\times10^{-4}$ \\
\hline
		$2$ & $-0.035\le x\le 0.035$ & $1.4\times10^{-4}$ \\
\hline
	\end{tabular}
 	\caption{Under different values of $r_0$, the minimum mertic $g_{tt}$ and its corresponding $x$ coordinate.}
	\label{tab:t2}
\end{table}

\begin{figure}
  \begin{center}
\subfigure{\includegraphics[width=0.49\textwidth]{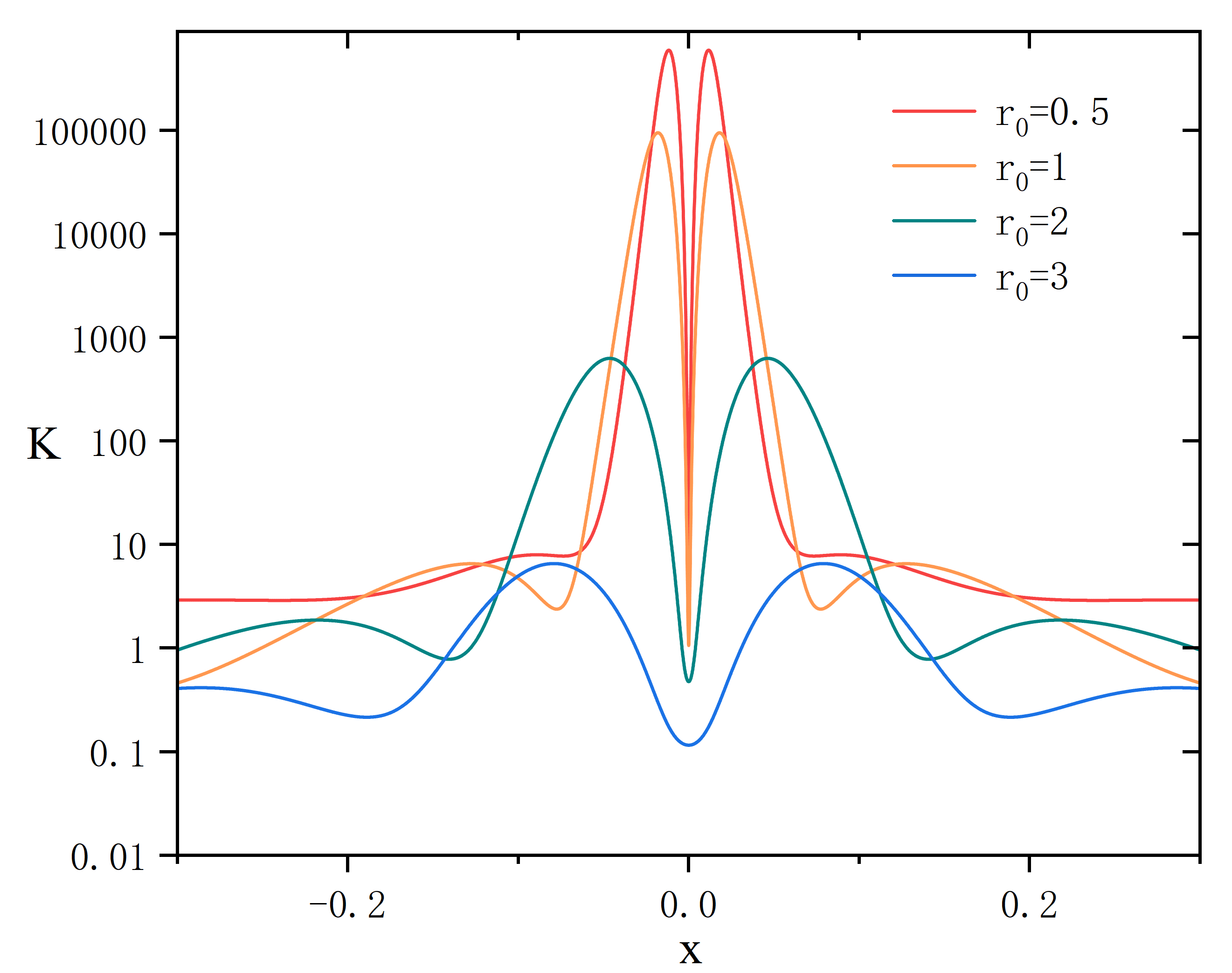}}
  \end{center}
\caption{The distribution of the Kretschmann scalar for the symmetric type II.}
\label{k2}
\end{figure}

\subsection{Asymmetric results}
In the asymmetric solution, besides analyzing the result of the field functions, ADM mass $M$ and Noether charge $Q$. We also explore the transformation between symmetric and asymmetric solutions.

\begin{figure}
  \begin{center}
\subfigure{\includegraphics[width=0.49\textwidth]{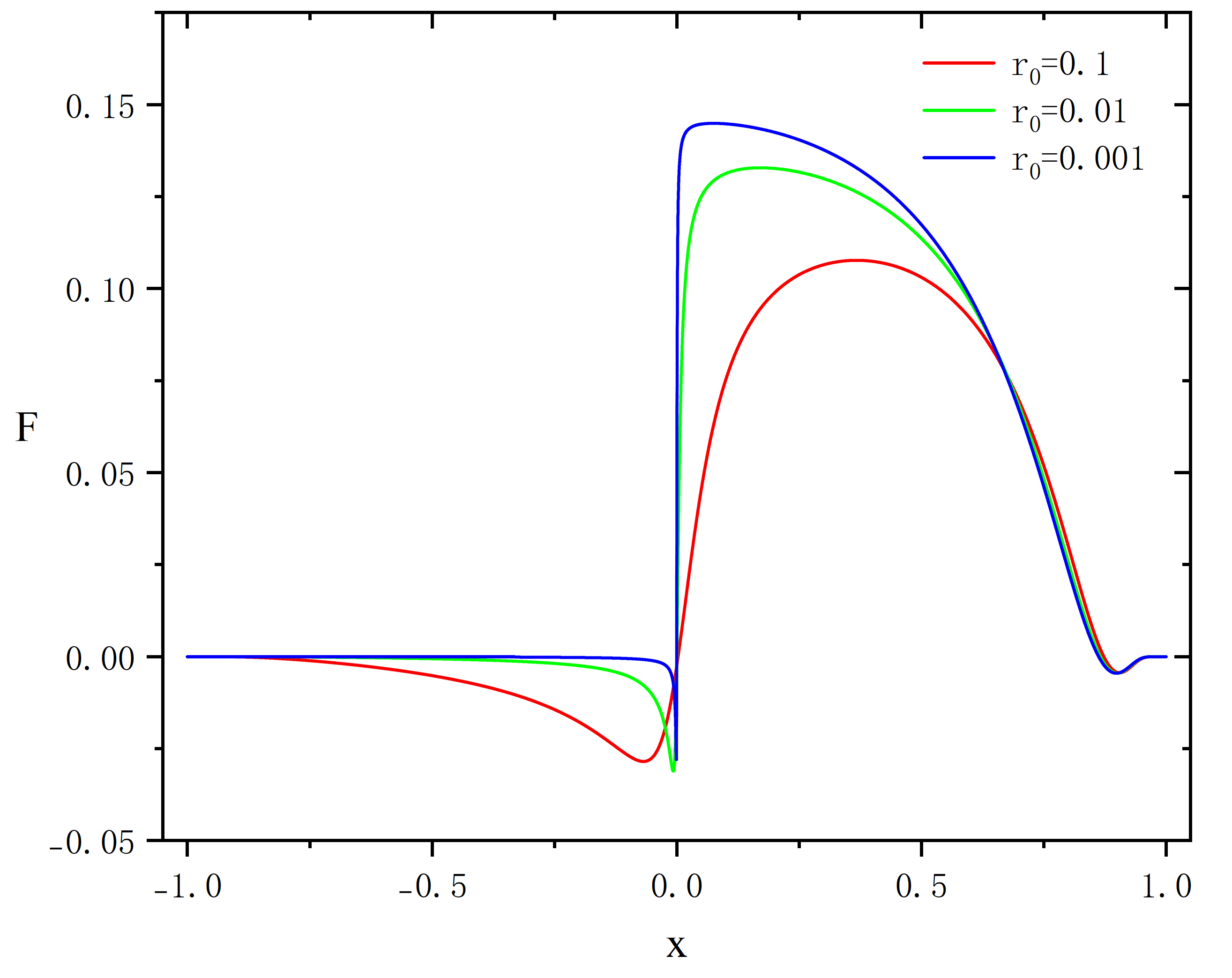}}
\subfigure{\includegraphics[width=0.48\textwidth]{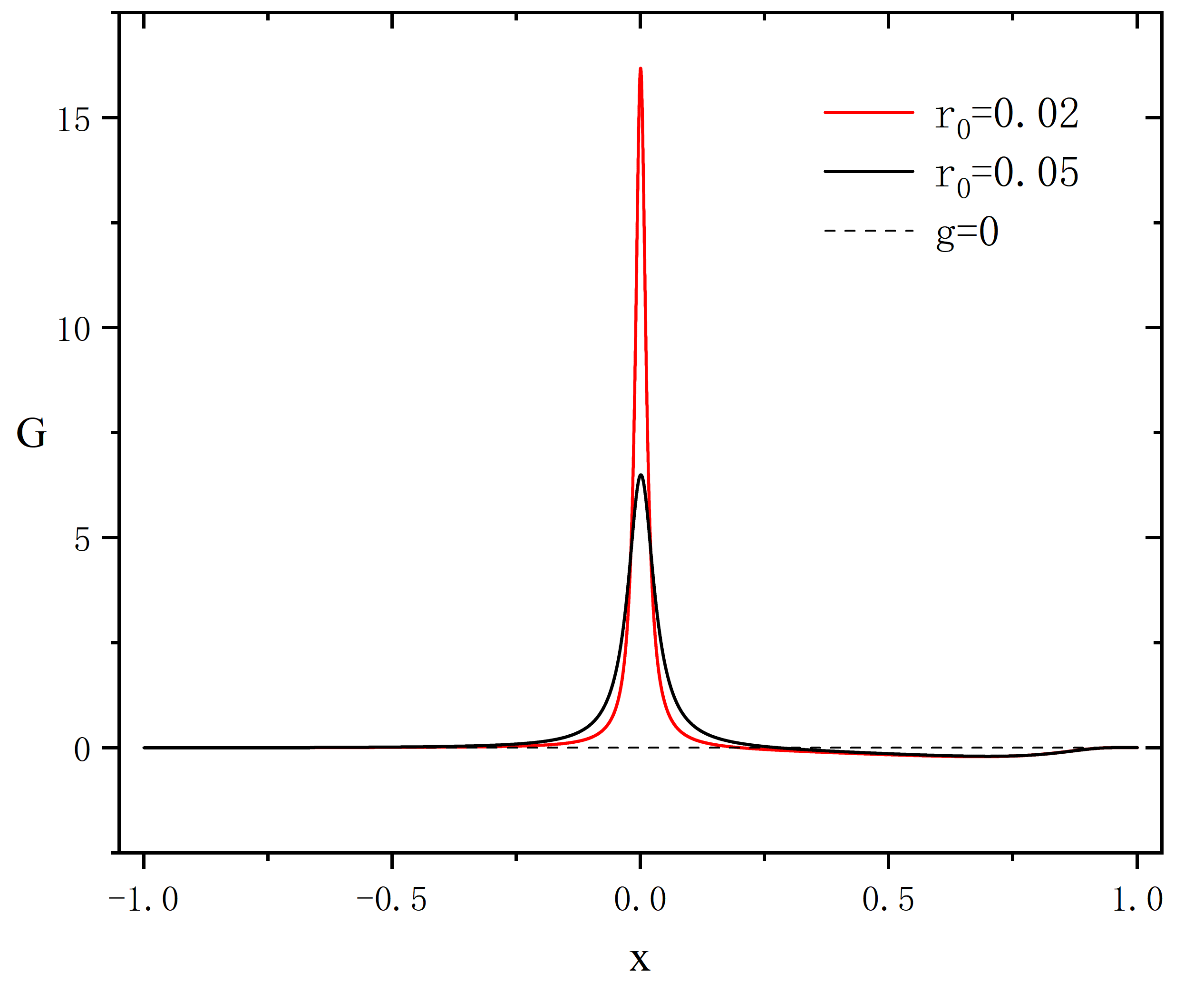}}
  \end{center}
\caption{In the case of asymmetry, when $r_{0}$ takes different values, the function images of field functions $F$ and $G$ with respent to $x$. }
\label{fg3}
\end{figure}

In Fig. \ref{fg3}, we show the results of the field functions $F$ and $G$, where the function $F$ has two nodes and $G$ has one node. We find that the function $F$ is mainly distributed in the interval of $x>0$, and when $r_{0}$ decreases, the part of the function in the interval of $x<0$ gets close to zero. This distribution of the $F$ field determines the property of the ADM mass and Noether charge, which we will discuss below.

\begin{figure}
  \begin{center}
\subfigure{\includegraphics[width=0.5\textwidth]{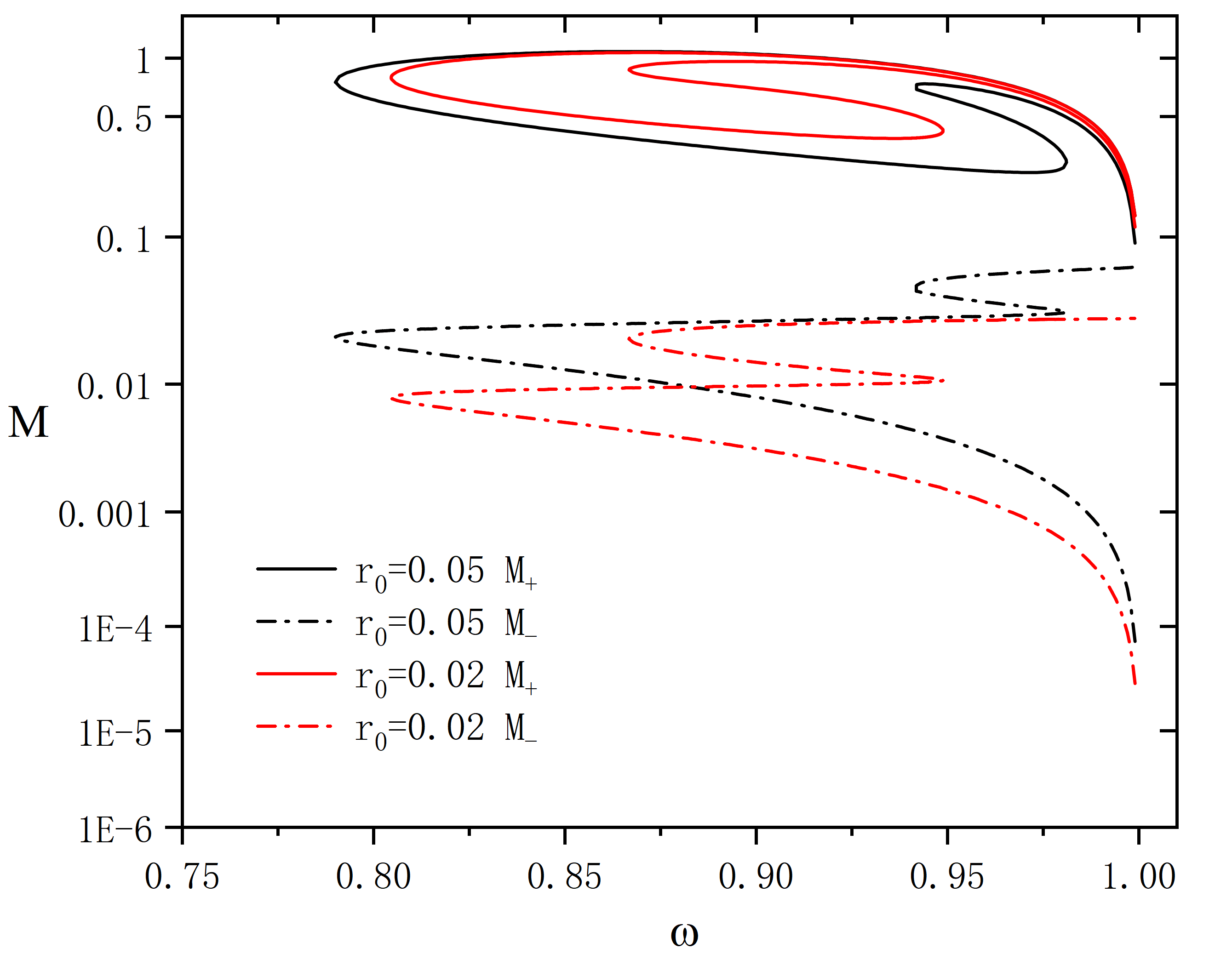}}
\subfigure{\includegraphics[width=0.48\textwidth]{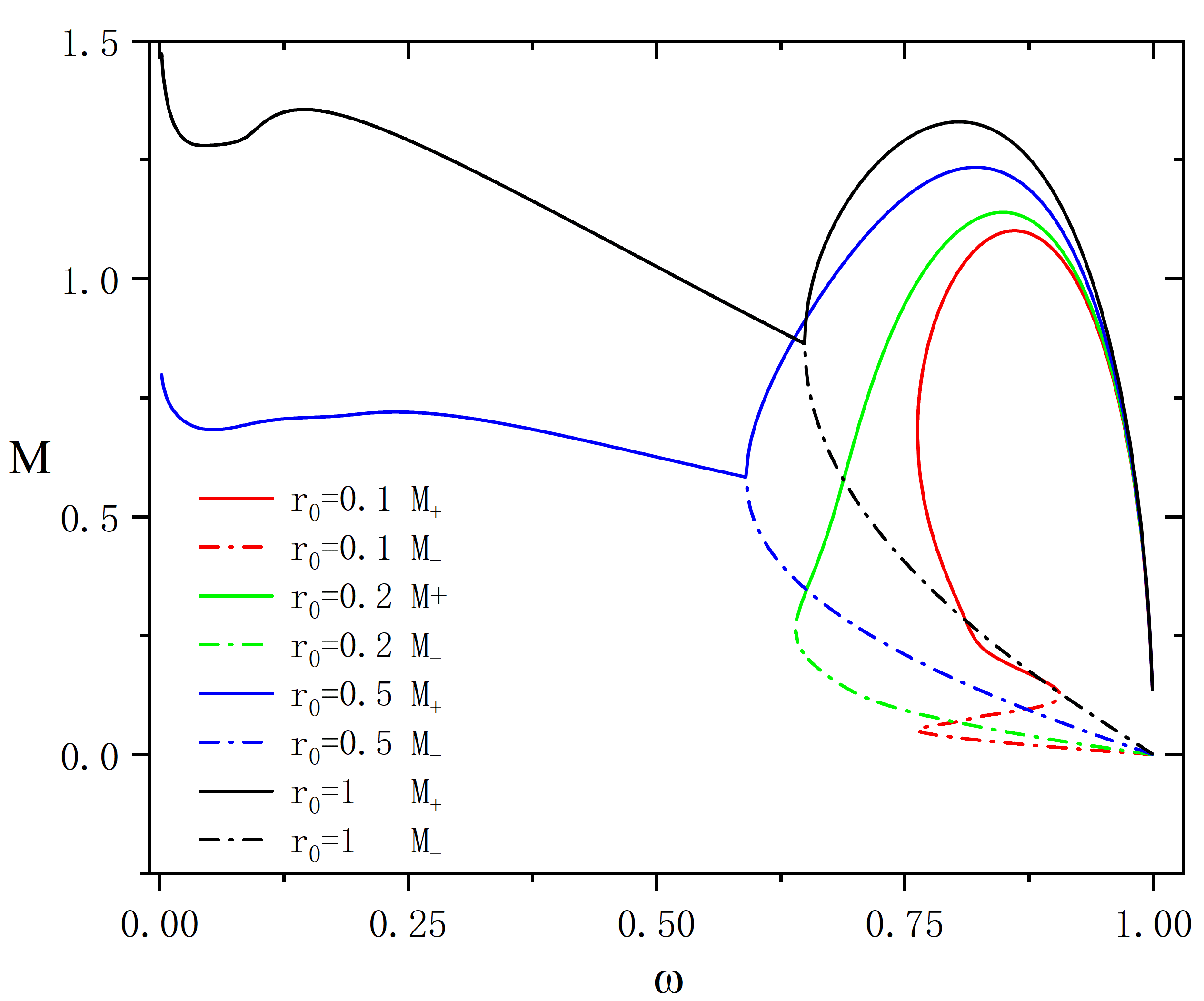}}
\subfigure{\includegraphics[width=0.5\textwidth]{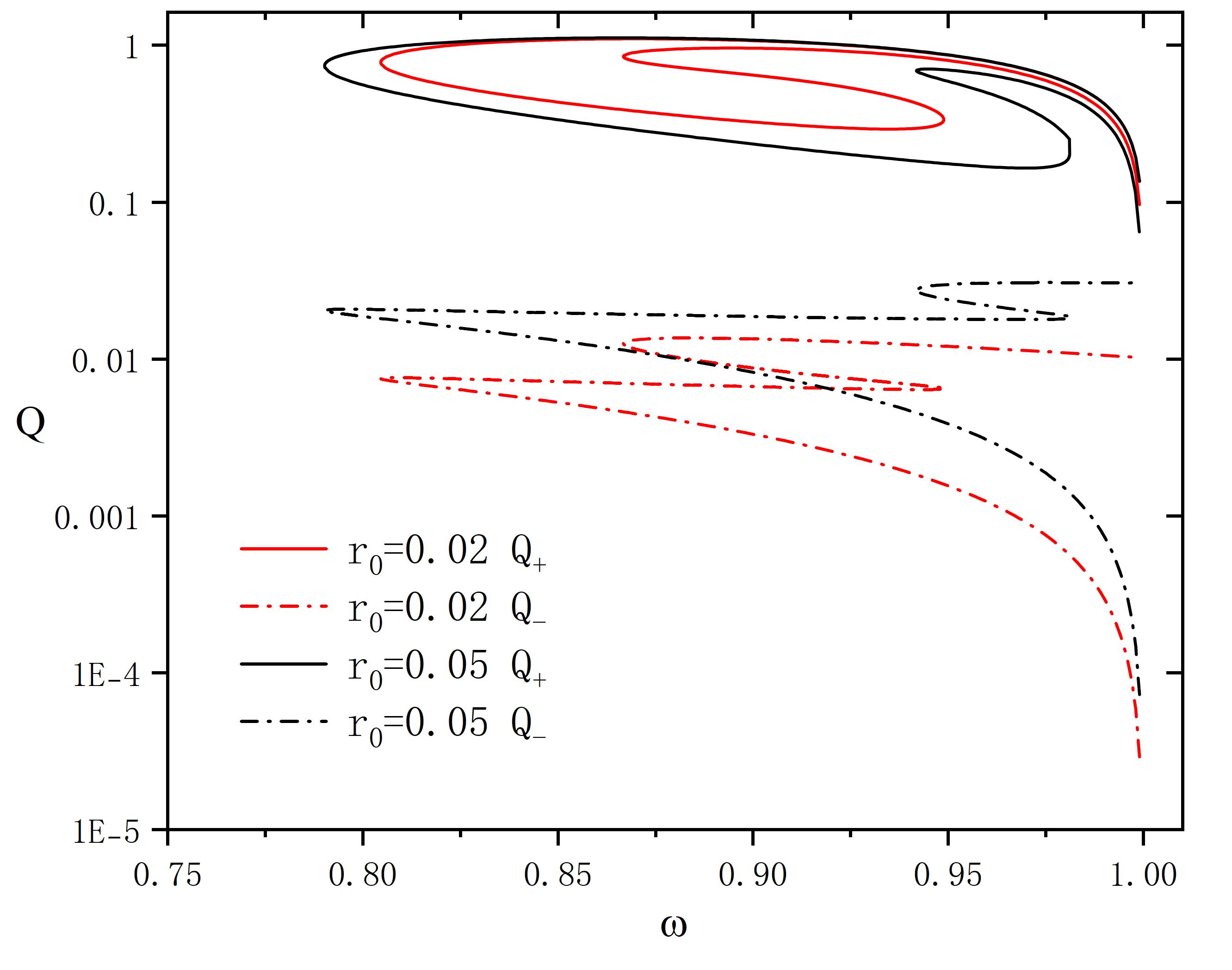}}
\subfigure{\includegraphics[width=0.49\textwidth]{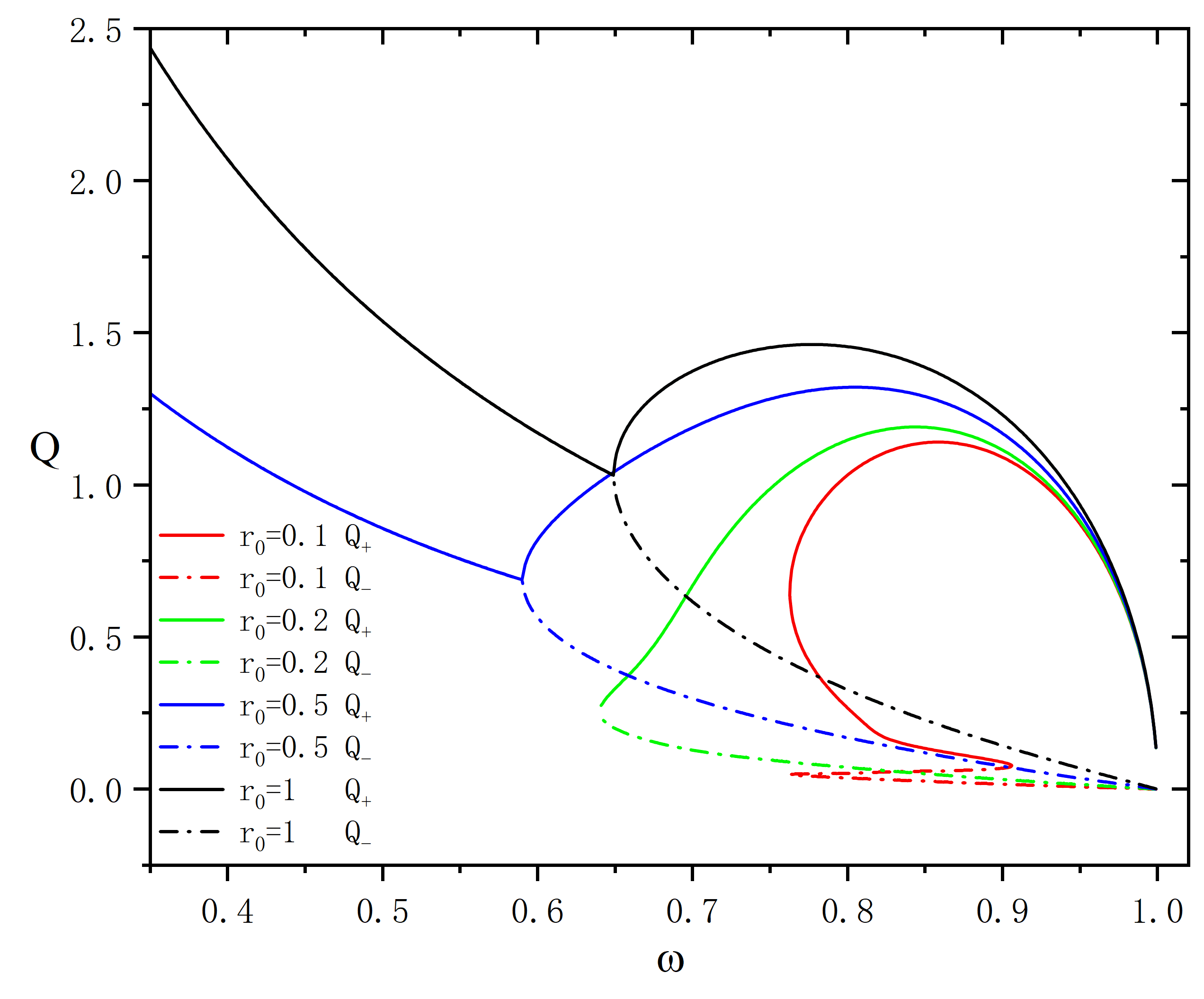}}
  \end{center}
\caption{In the case of asymmetry, the mass $M$ and the charge $Q$ as functions of frequency $\omega$ for different values of $r_{0}$.}
\label{mq3}
\end{figure}

\begin{figure}
  \begin{center}
\subfigure{\includegraphics[width=0.51\textwidth]{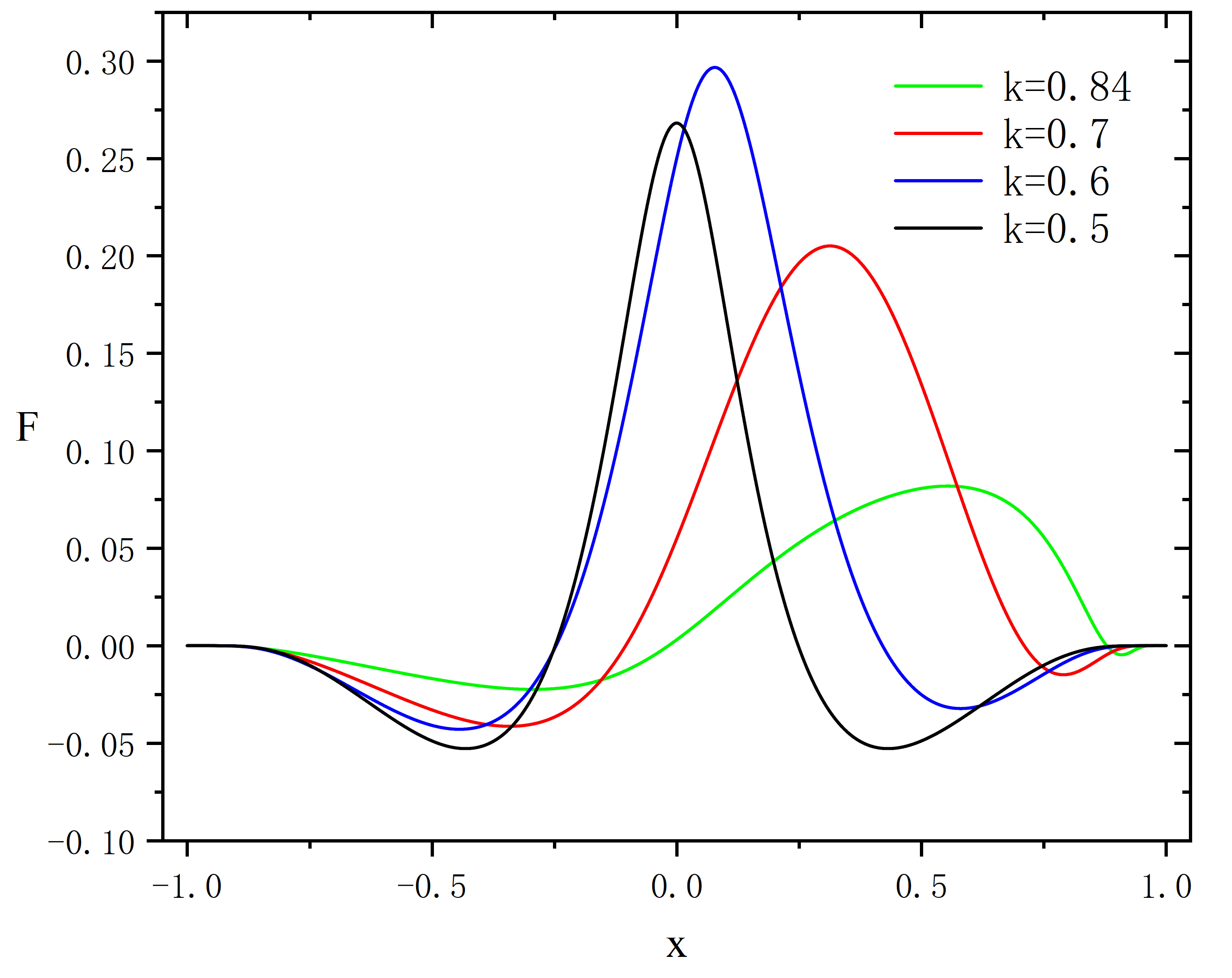}}
\subfigure{\includegraphics[width=0.48\textwidth]{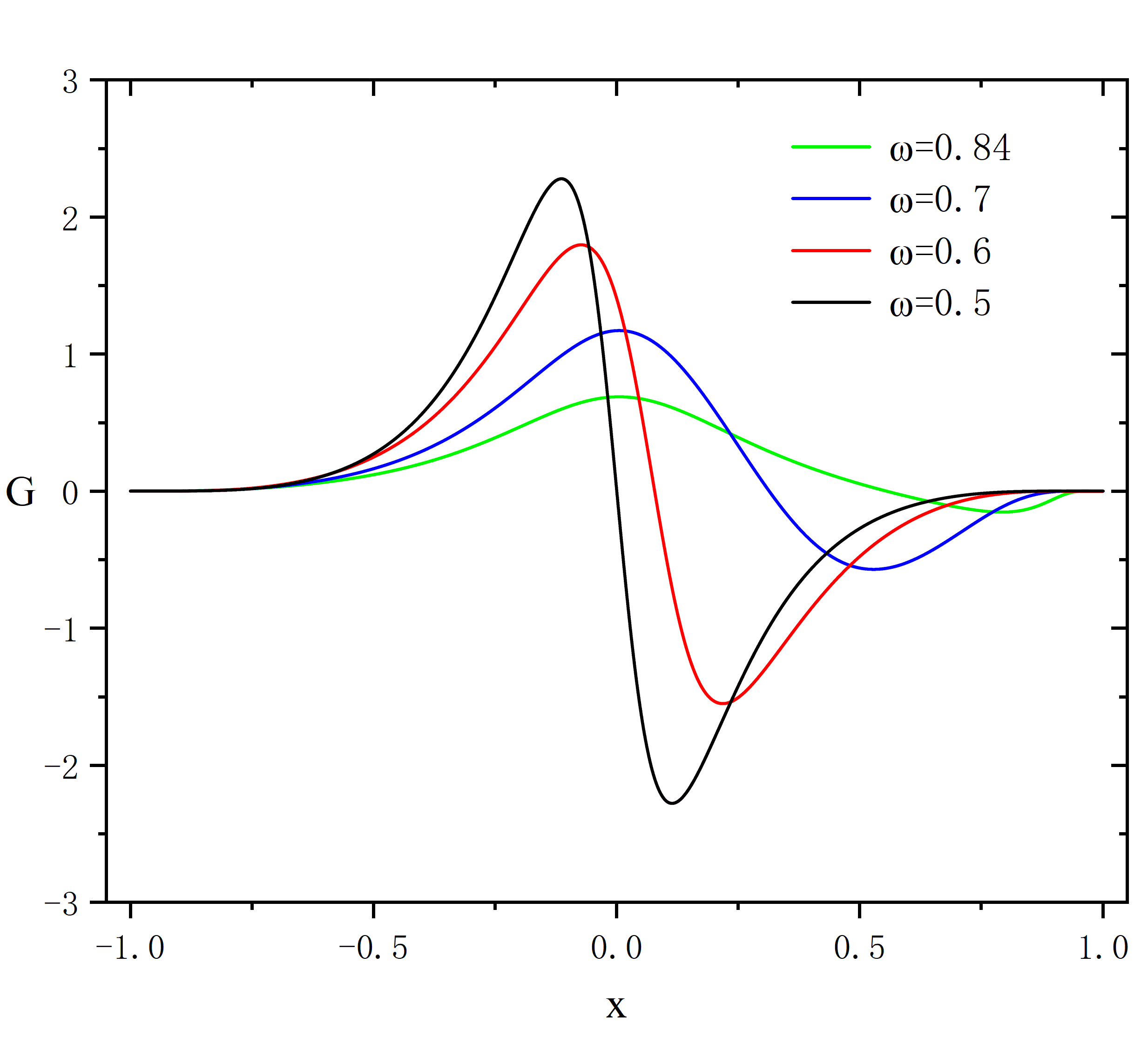}}
  \end{center}
\caption{Proca field $F$ and $G$ as functions of $x$ for $\omega=0.84, 0.7, 0.6, 0.5$ for $r_{0}=0.5$.}
\label{fg33}
\end{figure}

In Fig. \ref{mq3}, we show the ADM mass and Noether charge in the asymmetric solution, with $M_{+}$, $M_{-}$ representing the ADM mass at $x=+1,-1$, respectively, with $Q_{+}$, $Q_{-}$ represents the Noether charge $Q$ with the integration interval $x\in(0,+1)$,$x\in(-1,0)$. Similar to Fig. \ref{mq1} and Fig. \ref{mq2}, when $r_{0}$ is small, the solution is a multi-branch solution, and when $r_{0}$ is large, the solution has only one branch.

An interesting phenomenon is that when $r_{0}$ is small, the magnitudes of $M_{+}$ and $M_{-}$ (or $Q_{+}$ and $Q_{-}$) can differ by up to $10^{4}$, which is determined by the distribution of the field function $F$ in Fig. \ref{fg3}. When $r_{0}$ is larger, the curves of $M_{+}$ and $M_{-}$ will gradually merge together, the same applies to $Q_{+}$ and $Q_{-}$. This phenomenon is explained in Fig. \ref{fg33} where we show the change of the field functions with frequency for $r_{0}=0.5$. It can be found that when $\omega$ becomes smaller, the asymmetric solution becomes a symmetric solution. Such results indicate that under certain parameter settings, asymmetric solutions do not exist. It can be concluded that when the value of $r_{0}$ is large, as the frequency changes, the asymmetry decreases and the symmetry increases, eventually transforming into a symmetric solution. But when the value of $r_{0}$ is small, the asymmetric solution cannot be converted into a symmetric solution.

\begin{figure}
  \begin{center}
\subfigure{\includegraphics[width=0.49\textwidth]{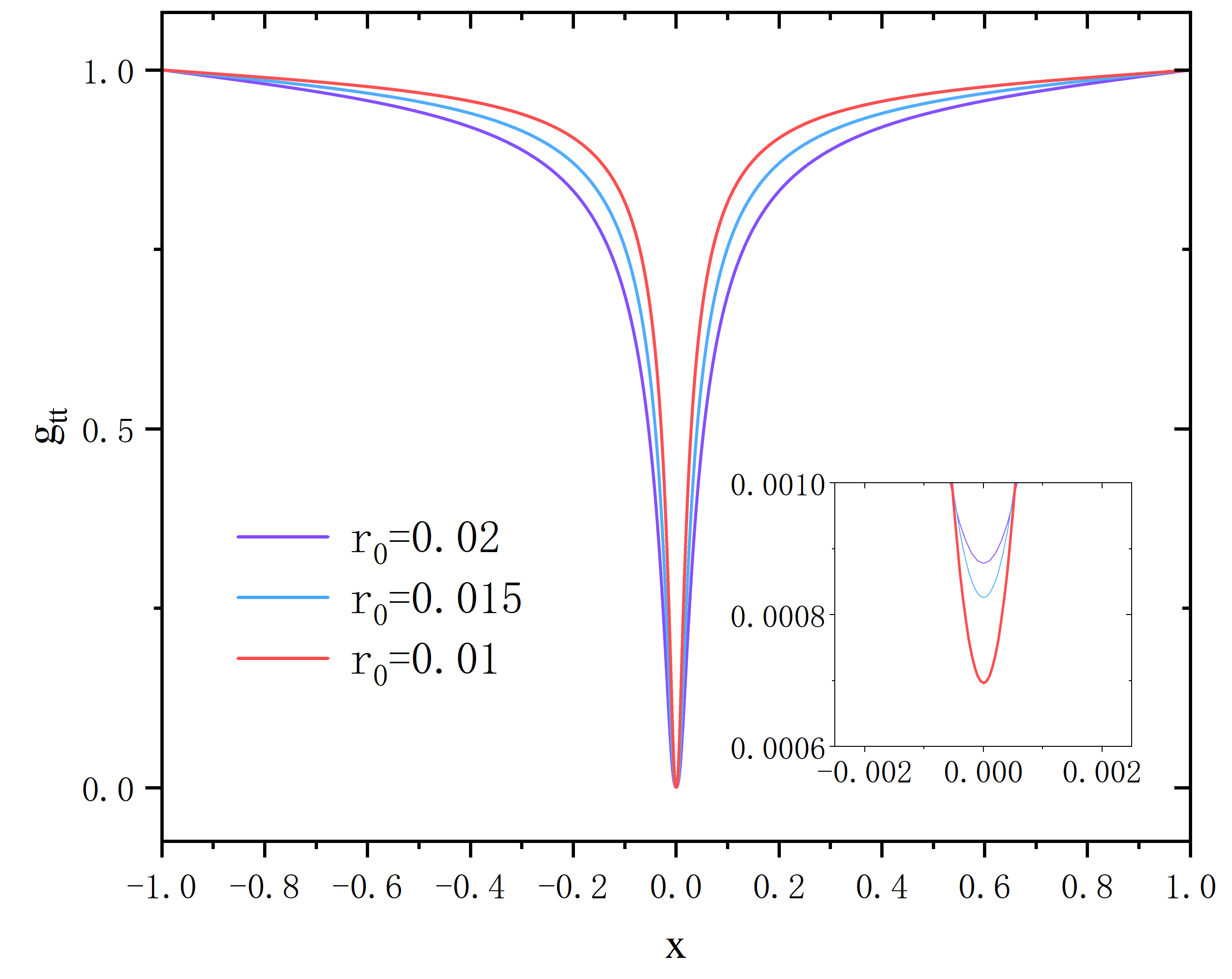}}
\subfigure{\includegraphics[width=0.49\textwidth]{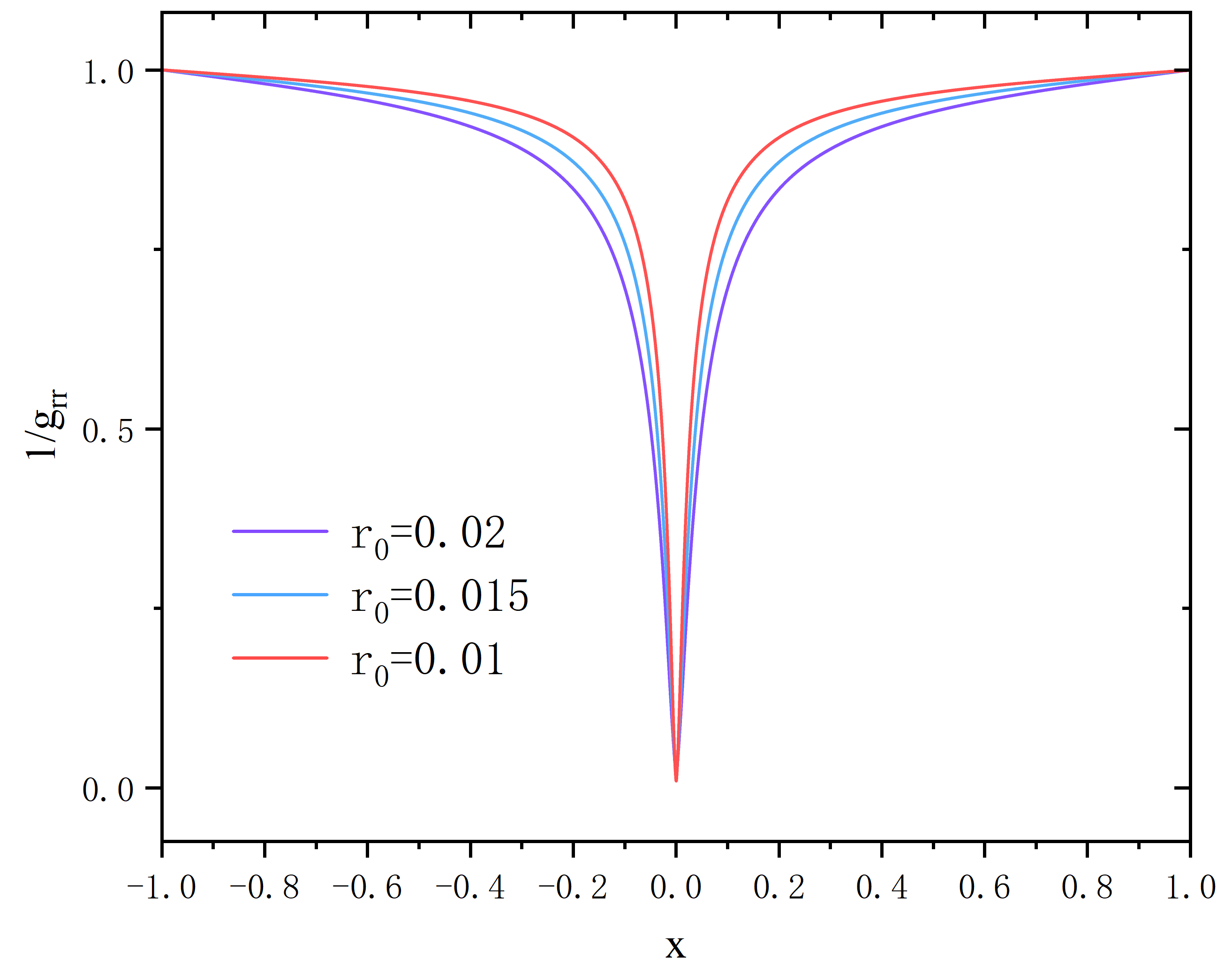}}
  \end{center}
\caption{$g_{tt}$ and $1/g_{rr}$ as functions of $x$ for the asymmetric solution, the frequency $\omega$ fixed in 0.9999. }
\label{gtr3}
\end{figure}

    	\begin{table}[ht] 
	\centering 
     \setlength{\tabcolsep}{16mm}
	\begin{tabular}{|c||c|c|}
\hline
		$r_0$ & $\omega$ & $g_{tt}(min)$ \\
\hline
		$0.02$ & $0.9999$ & $8.8\times10^{-4}$  \\
\hline
		$0.015$ & $0.9999$ & $8.3\times10^{-4}$ \\
\hline
		$0.01$ & $0.9999$ & $7.0\times10^{-4}$ \\
\hline
	\end{tabular}
 	\caption{Under different values of $r_0$, the minimum mertic $g_{tt}$ in the specific frequency.}
	\label{tab:t3}
\end{table}

\begin{figure}
  \begin{center}
\subfigure{\includegraphics[width=0.51\textwidth]{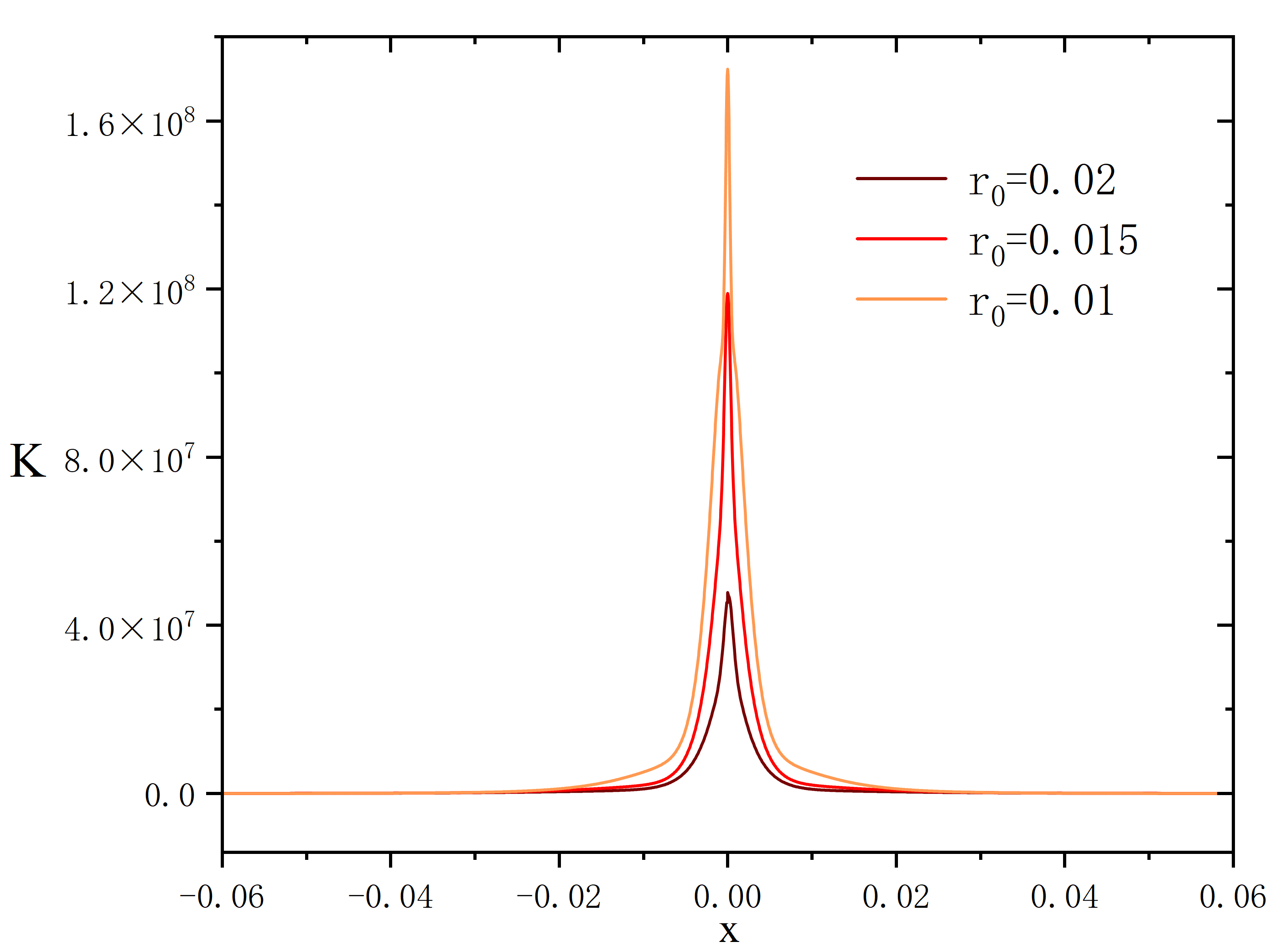}}
  \end{center}
\caption{The distribution of the Kretschmann scalar for the asymmetric solution, the frequency $\omega$ fixed in 0.9999.}
\label{fank}
\end{figure}

In addition, we consider whether in the case of asymmetric solutions, extreme approximations to black holes will also occur like symmetric solutions. We find that when $r_{0}$ is very small, the field function tends to diverge when $x$ is close to zero, and the smaller $r_{0}$ is, the more the function diverges. However, the general material field does not diverge, so we further analyzed the metric function when $r_{0}$ is small. As shown in Fig. \ref{gtr3} is the image of the metric function $g_{tt}$ and $1/g_{rr}$ when $r_{0}$ is 0.02, 0.015, 0.01. It can be found that when $x=0$, $g_{tt}$ and $1/g_{rr}$ reach the minimum value, and the minimum value is very close to zero. The smaller the size of the throats, the closer the minimum values of $g_{tt}$ and $1/g_{rr}$ are to zero, the specific results are shown in Tab. \ref{tab:t3}. Furthermore, we calculate the Kretschmann scalar under the same parameters in Fig. \ref{fank}. It can be seen that the Kretschmann scalar reaches an astonishing order of magnitude and diverges. This means the emergence of singularity and the wormhole is untraversable. In particular, the position $x$ where K diverges coincides with the position of the minimum point of $g_{tt}$, which means a curvature singularity appears at the event horizon. Therefore, we believe that at position $x=0$, there is an extremely approximate black hole solution in asymmetric solutions.  

\subsection{The $D$ charge}
\begin{figure}
  \begin{center}
\subfigure[~]{\label{ddd1}
\includegraphics[width=0.47\textwidth]{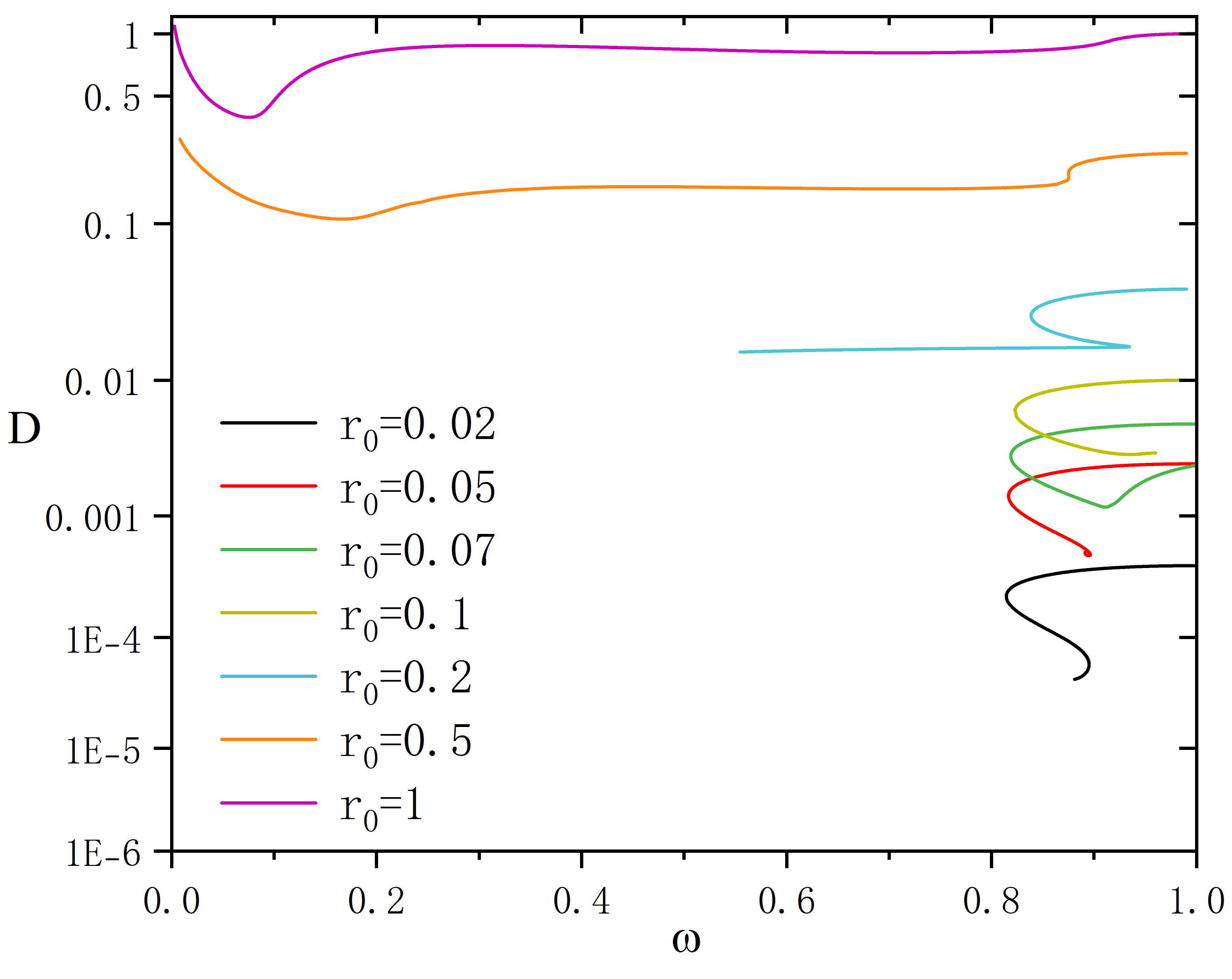}}
\subfigure[~]{\label{ddd2}
\includegraphics[width=0.47\textwidth]{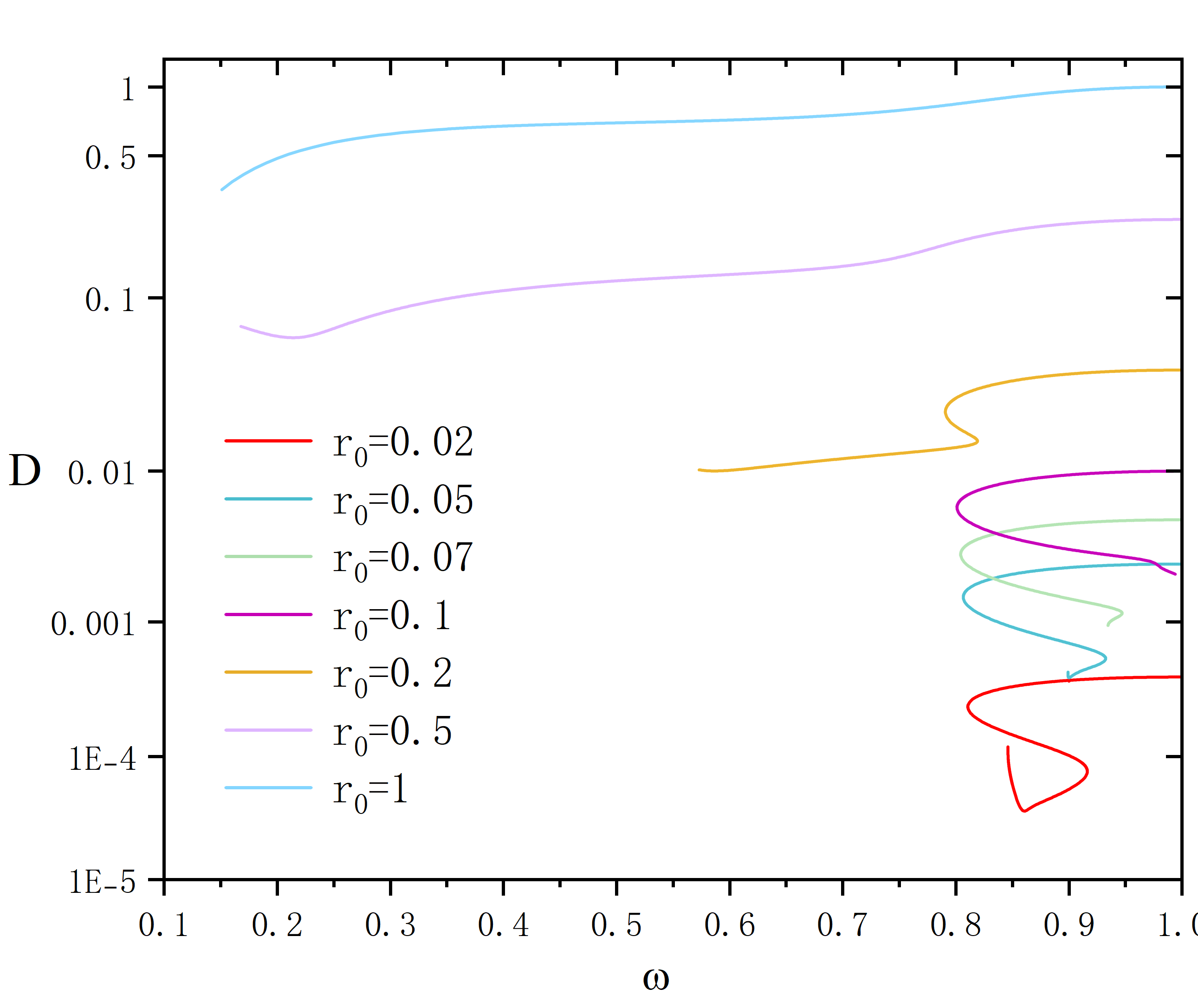}}
\subfigure[~]{\label{ddd3}
\includegraphics[width=0.47\textwidth]{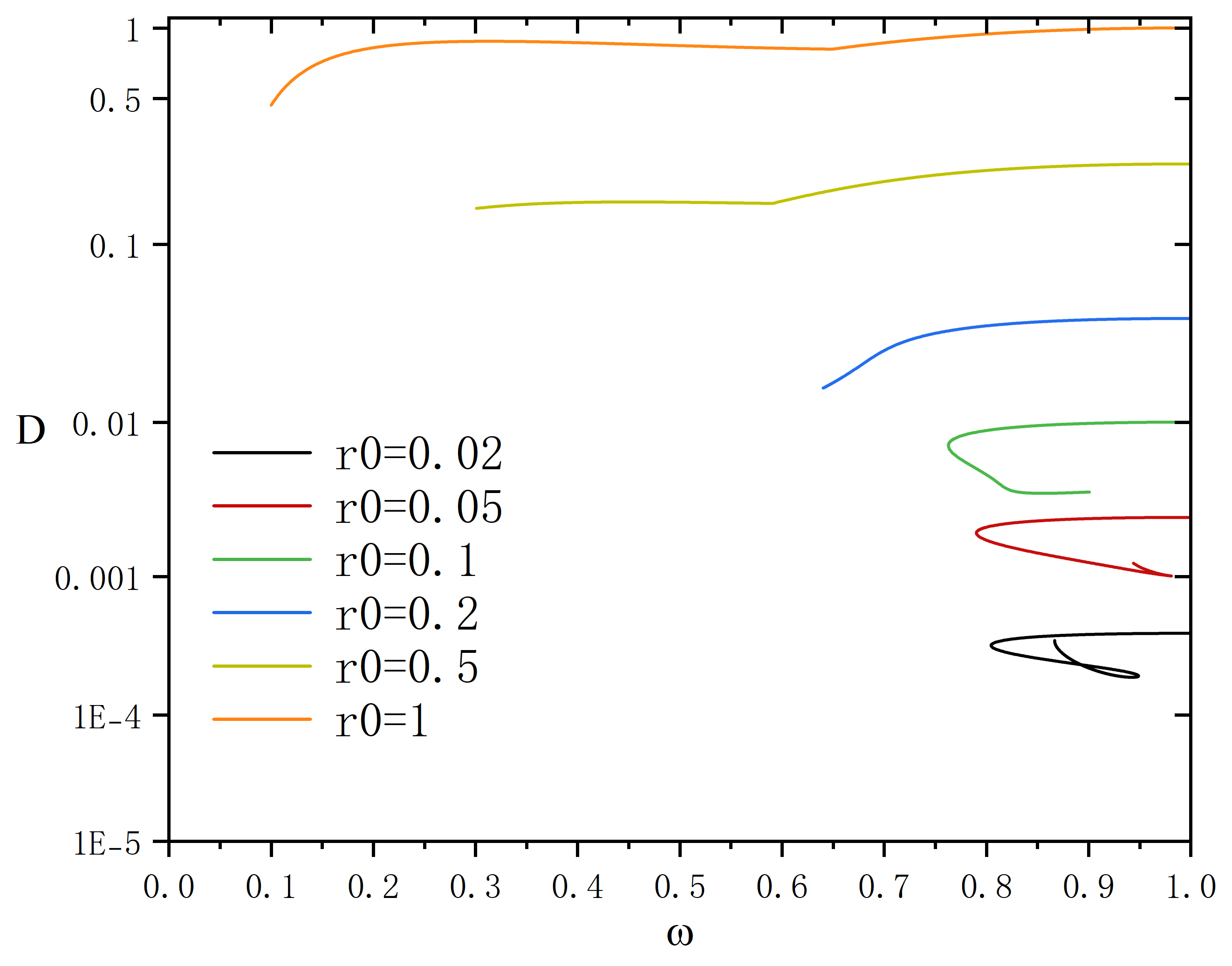}}
  \end{center}
\caption{The scalar charge $D$ of the phantom field as a function of the frequency $\omega$ with several values of the throat size $r_{0}$ for the symmetric type I (a), symmetric type II (b) and the asymmetric solution (c).  }
\label{ddd}
\end{figure}

In Fig. \ref{ddd} we show the relationship between the scalar charge $D$ and the frequency $\omega$ for different values of the throat size $r_{0}$ in the symmetric and asymmetric solutions. On the one hand, the scalar charge $D$ can be used to verify the accuracy of numerical calculations. Specifically, we calculate the value of $D$ at different $x$ positions and observe the fit of the function curves. We only retain the results with a difference of less than $10^{-4}$. On the other hand, the value of $D$ can reflect the amount of phantom matter. From the graphs, we can see that with the increase of $r_{0}$, the scalar charge also increases, which means that more phantom matter is required.

\subsection{Wormhole geometries}

Finally, we study the geometric properties of the wormhole. The method is to fix the $t$ and $\theta$ to obtain a two-dimensional hypersurface, which allows us to better observe the shape of the wormhole and the number of throats. Then we can embed the two-dimensional hypersurface into a three-dimensional Euclidean space, The resulting embedding map visualizes the wormhole geometry. This technique allows us to better understand the topology and properties of wormhole solutions.  

The specific steps are as follows. Fix the time $t$ and $\theta$($\theta=\pi/2$) and then use the cylindrical coordinates $(\rho ,\varphi,z )$, the line element can be written as
\begin{equation}
\begin{split}
ds^{2}&=Be^{-A} dr^{2}+Be^{-A} hd \varphi^{2}
\\&=d \rho^{2}+d z^{2}+\rho^{2} d \varphi^{2} .
\end{split}
\end{equation}

By comparing the metric forms in spherical coordinates and cylindrical coordinates, we can obtain expressions for $\rho$ and $z$
\begin{equation}
\rho (r)=\sqrt{B(r)e^{-A(r)}h(r)} ,  \ z(r)=\pm\int \sqrt{B(r)e^{-A(r)}-\left ( \frac{d\rho}{dr}  \right )^{2} } dr.
\end{equation}

Here $\rho$ corresponds to the circumferential radius, which corresponds to the radius of a circle located in the equatorial plane and having a constant coordinate $r$. The function $\rho$ has extreme points, where the first derivative is zero. When the second derivative of the extreme point is greater than zero, we call the point a throat, which corresponds to a minimal surface. When the second derivative of the extreme point is less than zero, we call the point the equator, which corresponds to the maximum surface.

\begin{figure}[!htbp]
  \begin{center}
\subfigure[~]{\label{figwh1}
\includegraphics[width=0.46\textwidth]{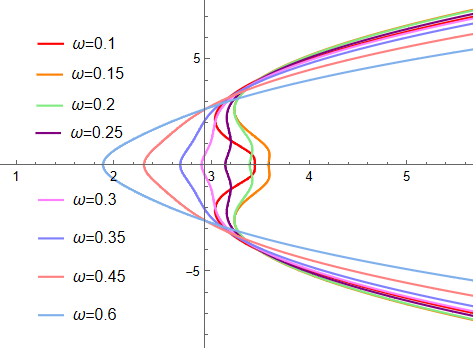}}
\subfigure[~]{\label{figcd1}
\includegraphics[width=0.46\textwidth]{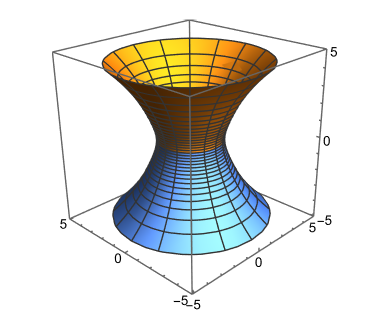}}
\subfigure[~]{\label{figbl1}
\includegraphics[width=0.46\textwidth]{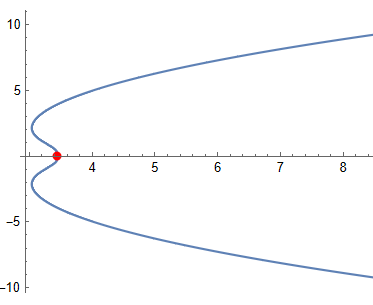}}
\subfigure[~]{\label{figcd2}
\includegraphics[width=0.46\textwidth]{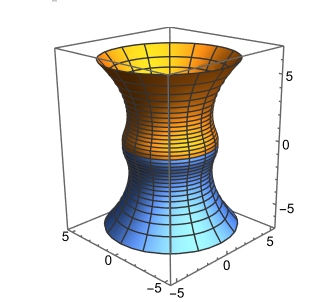}}
  \end{center}
\caption{These four pictures are for the symmetric type I. (a): Two-dimensional view of the isometric embedding of the equatorial plane for different $\omega$ with throat parameter  $r_{0} = 1$. (b): Isometric embeddings of the solution with throat parameter $r_{0} = 1$ and $\omega = 0.6$. (c) and (d): Corresponding two-dimensional view of the isometric embedding of the equatorial plane wormhole embedding diagram when $\omega = 0.1$, The red dot in (c) represents the position of the extreme approximate black hole.}
\label{chong1}
\end{figure}

\begin{figure}[!htbp]
  \begin{center}
\subfigure[~]{\label{figwh2}
\includegraphics[width=0.47\textwidth]{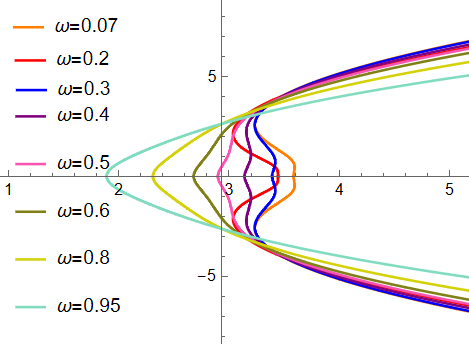}}
\subfigure[~]{\label{figcd3}
\includegraphics[width=0.46\textwidth]{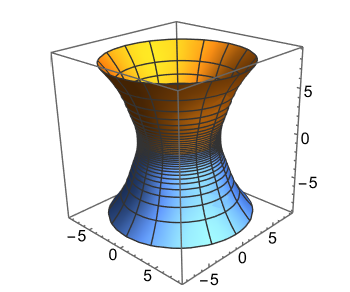}}
\subfigure[~]{\label{figbl2}
\includegraphics[width=0.46\textwidth]{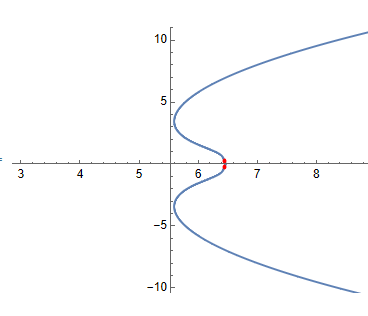}}
\subfigure[~]{\label{figcd4}
\includegraphics[width=0.47\textwidth]{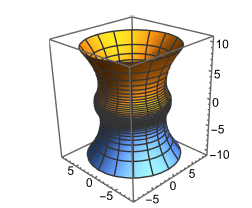}}
  \end{center}
\caption{These four pictures are for the symmetric type II. (a) : Two-dimensional view of the isometric embedding of the equatorial plane for different $\omega$ with throat parameter  $r_{0} = 2$. (b) : Isometric embeddings of the solution with throat parameter $r_{0} = 2$ and $\omega = 0.5$. (c) and (d): Corresponding two-dimensional view of the isometric embedding of the equatorial plane wormhole embedding diagram when $\omega = 0.07$, The red dots in (c) represents the position of the two extreme approximate black hole with symmetry about the throat.}
\label{chong2}
\end{figure}

\begin{figure}[!htbp]
  \begin{center}
\subfigure[~]{\label{figwh3}
\includegraphics[width=0.47\textwidth]{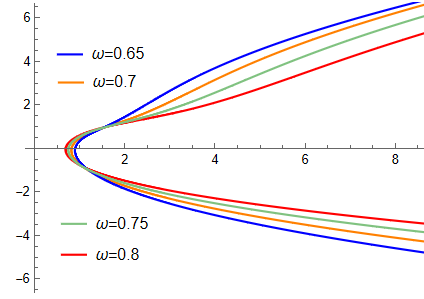}}
\subfigure[~]{\label{figcd5}
\includegraphics[width=0.49\textwidth]{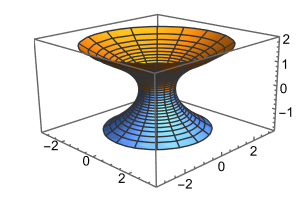}}
\subfigure[~]{\label{figbl3}
\includegraphics[width=0.47\textwidth]{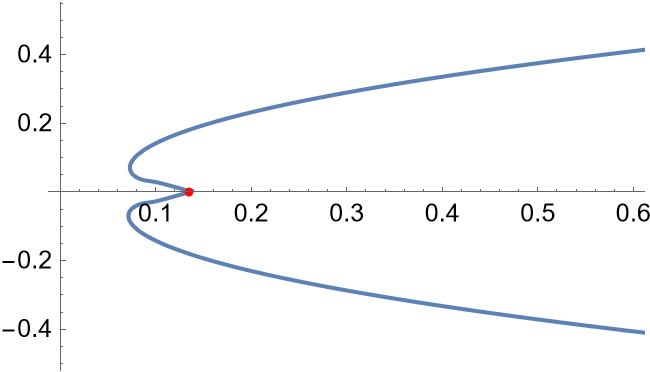}}
\subfigure[~]{\label{figcd6}
\includegraphics[width=0.46\textwidth]{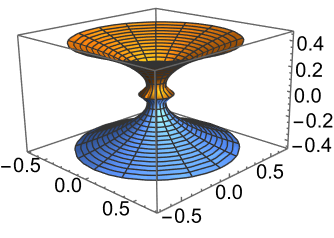}}
  \end{center}
\caption{These four pictures are for the asymmetric solution. (a) : Two-dimensional view of the isometric embedding of the equatorial plane for different $\omega$ with throat parameter  $r_{0} = 0.5$. (b) : Isometric embeddings of the solution with throat parameter $r_{0} = 0.5$ and $\omega = 0.75$. (c) and (d): Corresponding two-dimensional view of the isometric embedding of the equatorial plane wormhole embedding diagram with throat parameter $r_{0} = 0.02$ and $\omega = 0.99$, The red dot in (c) represents the position of the extreme approximate black hole.}
\label{chong3}
\end{figure}

To illustrate the geometric properties of throats, we show two--dimensional views of the isometric embedding of the equatorial plane and the wormhole geometry with extremely approximate black holes. For the two types of symmetric solutions, we choose throat parameters $r_{0}=1$ and $r_{0}=2$ respectively to draw two-dimensional equatorial planes at different frequencies. Fig. \ref{chong1} and Fig. \ref{chong2} depict that at high frequencies, the wormhole has only one throat with a small radius. As the frequency rises, the radius of the throat also increases. At near $\omega=0.25$, two additional throats start to emerge, and the wormhole then has three throats. With further increases in frequency, the radius of the three throats will first increase and then decrease. In particular, the middle throat will become less and less visible, although it always exists, but tends to disappear. 

In Fig. \ref{chong3}, we present two--dimensional views of the isometric embedding of the equatorial plane of asymmetric solutions. We have selected an additional set of parameter plots to reflect the emergence of extreme approximate black holes. In addition, there is a tendency from asymmetry to more asymmetry as the frequency changes.

\section{CONCLUSION}\label{sec5}

In this paper, we investigate a new type of traversable wormhole model by constructing a Proca star with a phantom field, obtaining two types of symmetric solutions and one asymmetric solution. The ADM mass, Noether charge, and phantom field scalar charge embody the properties of the solution, and the embedded diagram vividly illustrates the geometric characteristics of the wormhole. 

For two types of symmetric solutions, we analyzed the influence of parameter $r_{0}$ on the shape of the resulting curve. It is found that when $r_{0}$ is very small for the symmetric type I solution, the curves of the ADM mass and Noether charge initially exhibit a loop structure. As $r_{0}$ increases, the multi-branch curve evolves into a single-branch curve. Similarly, the resulting curve of the type II solution initially displays a spiral structure, then forms a loop as $r_{0}$ increases, and eventually becomes a single-branch curve. In particular, the symmetric type I solution can degenerate into two Proca stars in the asymptotically flat region when $r_{0}$ is zero, while the other symmetric case and the asymmetric case cannot degenerate into Proca stars. In the asymmetric case, when the parameter $r_{0}$ takes a small value, the difference between $M_{+}$ and $M_{-}$, $Q_{+}$ and $Q_{-}$ is large. As $r_{0}$ increases, the curves slowly intersect with the change of frequency, indicating a transition from asymmetry to symmetry.

Notably, we find that there are extreme approximate black hole solutions in this static model. Particularly, for the symmetric type II, there is an extremely approximate black hole on both sides of the wormhole throat. This configuration gives rise to a black hole-wormhole-black hole combination, a model we have always been interested in ER=EPR theory. In previous research, B. Kain discovered during the numerical evolution of wormholes that EDM wormholes will become untraversable over time, resulting in the appearance of black holes on both sides of the wormhole \cite{Kain:2023ann,Kain:2023ore}. Our study provides, for the first time, a statically stable model of black holes appearing on both sides of a wormhole offering robust support for the ER=EPR theory.

Our research work could make some interesting extensions in the future. For example, traversable wormholes without exotic matter can be constructed under modified gravity theory. Therefore, it is interesting to study the existence of such solutions under modified gravity.  Additionally, we know that Einstein-Dirac-Maxwell wormholes are also traversable wormholes, so we can consider adding the Maxwell field to our model to explore its properties. More importantly, extreme approximate black holes appear in our model, so we will further explore the connection between black holes and wormholes.

\section{ACKNOWLEDGEMENTS}
Thanks to Burkhard Kleihaus for explaining the calculation of the Noether charge. This work is supported by the National Key Research and Development Program of China (Grant No. 2020YFC2201503) and the National Natural Science Foundation of China (Grant No. 12047501 and No. 12275110). Parts of computations were performed on the shared memory system at the Institute of Computational Physics and Complex Systems at Lanzhou University.

\end{document}